\documentclass[fleqn,usenatbib]{mnras}

\usepackage[T1]{fontenc}
\usepackage{ae,aecompl}
\usepackage{xtab}
\usepackage{courier}
\usepackage{multirow}
\usepackage{times}
\usepackage{upgreek}
\usepackage{graphicx}
\usepackage{epstopdf}
\usepackage{amssymb}
\usepackage{amsmath}
\usepackage[section]{placeins}

\usepackage{newtxtext,newtxmath}

\usepackage[T1]{fontenc}
\usepackage{ae,aecompl}

\usepackage{graphicx}	
\usepackage{amsmath}	
\usepackage{amssymb}	

\usepackage[usenames,dvipsnames]{xcolor}
\usepackage{soul}

 \newcommand{\cmss}{\,\mbox{$\mbox{cm}\,\mbox{s}^{-2}$}}    

\title[Mode identification in TESS observed sdBVs]{Mode identification in three pulsating hot subdwarfs observed with TESS satellite}
\author[S.K.\,Sahoo et al.]{S.\,K.\,Sahoo$^{1,2}$\thanks{E-mail: sumanta.kumar27@gmail.com},
A.\,S.\,Baran$^{1}$,
U.\,Heber$^{3}$,
J.\,Ostrowski$^{1}$,
S.\,Sanjayan$^{1,2}$,
R.\,Silvotti$^{4}$,
\newauthor
A.\,Irrgang$^{3}$,
M.\,Uzundag$^{5,6}$,
M.\,D.\,Reed$^{1,7}$,
K.\,A.\,Shoaf$^{7}$,
R.\,Raddi$^{3,8}$,
M.\,Vuckovic$^{5}$,
\newauthor
H.\,Ghasemi$^{9}$,
W.\,Zong$^{10}$
and K.\,J.\,Bell$^{11,12}$
\\
$^{1}$ARDASTELLA Research Group, Institute of Physics, Pedagogical University of Krakow, ul. Podchor\c{a}\.zych 2, 30-084 Krak\'ow, Poland\\
$^{2}$Nicolaus Copernicus Astronomical Centre of the Polish Academy of Sciences, ul. Bartycka 18, 00-716 Warsaw, Poland\\
$^{3}$Dr. Remeis-Sternwarte and ECAP, Astronomical Institute, University of Erlangen-N\"urnberg, Sternwartstr. 7, D-96049 Bamberg, Germany\\
$^{4}$INAF-Osservatorio Astrofisico di Torino, Strada dell'Osservatorio 20, I-10025 Pino Torinese, Italy\\
$^{5}$Instituto de Fisica y Astronomia, Facultad de Ciencias, Universidad de Valparaiso, Gran Bretana 1111, Playa Ancha 2360102 Valparaiso, Chile\\
$^{6}$European Southern Observatory, Alonso de Cord\'ova 3107, Santiago, Chile\\
$^{7}$Department of Physics, Astronomy, and Materials Science, Missouri State University, Springfield, MO 65897, USA\\
$^{8}$Universitat Politecnica de Catalunya, Departament de Fisica, c/ Esteve Terrades, 5, 08860 Castelldefels, Spain\\
$^{9}$Department of Physics, Institute for Advanced Studies in Basic Sciences (IASBS), Zanjan 45137-66731, Iran\\
$^{10}$Department of Astronomy, Beijing Normal University, Beijing 100875, PR China\\
$^{11}$DIRAC Institute, Department of Astronomy, University of Washington, Seattle, WA 98195-1580, USA\\
$^{12}$NSF Astronomy and Astrophysics Fellow}
\date{Accepted XXX. Received YYY; in original form ZZZ}

\pubyear{2020}

\begin{document}
\label{firstpage}
\pagerange{\pageref{firstpage}--\pageref{lastpage}}
\maketitle

\begin{abstract}
We report on the detection of pulsations of three pulsating subdwarf B stars observed by the TESS satellite and our results of mode identification in these stars based on an asymptotic period relation. SB\,459 (TIC\,067584818), SB\,815 (TIC\,169285097) and PG\,0342+026 (TIC\,457168745) have been monitored during single sectors resulting in 27\,days coverage. These datasets allowed for detecting, in each star, a few tens of frequencies, which we interpreted as stellar oscillations. We found no multiplets, though we partially constrained mode geometry by means of period spacing, which recently became a key tool in analyses of pulsating subdwarf B stars. Standard routine that we have used allowed us to select candidates for trapped modes that surely bear signatures of non-uniform chemical profile inside the stars. We have also done statistical analysis using collected spectroscopic and asteroseismic data of previously known subdwarf B stars along with our three stars. Making use of high precision trigonometric parallaxes from the Gaia mission and spectral energy distributions we converted atmospheric parameters to stellar ones. Radii, masses and luminosities are close to their canonical values for extreme horizontal branch stars. In particular, the stellar masses are close to the canonical one of 0.47\,M$_\odot$ for all three stars but uncertainties on the mass are large. The results of the analyses presented here will provide important constrains for asteroseismic modelling.
\end{abstract}




\section{Introduction}
Subdwarf B (sdB) stars are extreme horizontal branch stars, consist of a convective helium burning core, helium shell and a very thin (in mass) hydrogen envelope. The effective temperatures T$_{\rm eff}$ are in a range of 20,000 to 40,000\,K, which moved them blueward from the normal horizontal branch stars in the Hertzsprung-Russell diagram \citep{heber16}. The sdB stars are found in almost all stellar populations, field \citep{2004A&A...414..181A,2017MNRAS.467...68M} as well as open \citep{janusz93} and globular clusters \citep{moehler01,moni08}. The sdB stars have masses nearly 0.5\,M$_{\sun}$ and surface gravities, in a logarithmic scale, $\log{(g/\cmss)}$ of 5.0\,--\,5.8 \citep{heber16}, which means that they are compact in size (0.15\,--\,0.35\,R$_{\sun}$). They are considered to be one of the most ionizing sources of interstellar gas at high galactic latitudes \citep{deboer85}, and mostly responsible for the ultraviolet upturn phenomenon in early-type galaxies \citep{brown97}.

Due to a low mass of the hydrogen envelope (<$10^{-2}M_{\sun}$), sdB stars are not able to sustain two shell nuclear burning, skipping the Asymptotic Giant Branch, and heading directly to the white dwarf cooling track, right after the helium in the core is exhausted. The reason for the lack of a more massive hydrogen envelope is still a puzzle, though binarity is a natural explanation for a mass loss. This can explain sdBs in binaries.  A merger event can be invoked as a channel leading to formation of single sdBs \citep{han02}. According to \cite{charpinet18} who presented results on stellar rotation analysis, substellar companions may also be responsible for mass loss, while the objects are still observationally single. \cite{fontaine12} concluded that the mass distribution points at single star evolution, however in those cases a strong wind is necessary to remove the hydrogen envelope. No evidence for a strong wind has been reported thus far.

Discovery of pulsating sdB stars (hereafter: sdBV) by \citet{kilkenny97} has opened a way to understand their internal structure by using asteroseismological techniques \citep{charpinet97}. SdBV stars show pulsations in p-modes or g-modes, though a mix of the modes are recently commonly found in the so-called hybrid sdBV stars. Typical periods of p-mode pulsators are of the order of minutes, while periods of g-mode pulsators are of the order of hours \citep{heber16}.

In the field of sdBV stars, a significant improvement has been made during the last several years and a big credit goes to the Kepler and K2 missions due to their unprecedented data delivery. Asteroseismic analyses of Kepler-observed sdBV stars have revealed interesting and useful features. Rotationally split multiplets and asymptotic period sequences have never been easy to detect in ground-based data; Balloon\,090100001 being an exception \citep{baran09}. Multiplets allowed for identification of low degree ($\ell\leq2$) modes, although higher degrees ($3\leq\ell\leq8$) in the "intermediate region" of 400-700\,$\upmu$Hz have been also detected \citep{foster15,telting14,silvotti19}. An observed period spacing between consecutive overtones of g-modes of the same modal degree, ranges from 230 to 270\,seconds \citep{reed18b}. Asymptotic sequences often show a "hook" feature \citep[e.g.][]{baran12b} in \'echelle diagrams and occasionally include trapped modes \citep[e.g.][]{ostensen14}, which is likely the indication of a non-uniform chemical profile along a stellar radius \citep{charpinet00}. Pulsation models also predict low order p-mode overtones to be spaced in frequency by 800 -- 1,100\,$\upmu$Hz \citep{charpinet00}. Observations have yielded a mixture of results, with three sdBV stars in agreement \citep{baran09,foster15,reed19} and two other with much smaller spacings \citep{baran12a,reed19}.

The successor of Kepler and K2 missions, TESS \citep[Transiting Exoplanet Survey Satellite;][]{ricker14}, an all-sky survey, satellite has been launched on April 18, 2018. The main goal of the TESS mission is to detect exoplanets around nearby bright (down to about 15\,mag) stars by using the transit method. It provides data over a time span of two years by using its four CCD cameras with 24$\times$96 degree field of view, which is known as an individual sector. TESS will cover 26 sectors over 24\,months. The short cadence (SC) mode of 2\,min, allocated for a selected sample of targets, allows us to investigate the light variations of the pulsating subdwarf B stars, covering entire g-mode region and reaching up to the longest period p-modes.

This paper reports results of our work on three sdB stars monitored during the TESS mission and found to be light variable consistent with stellar pulsations. The targets SB\,459 (TIC\,067584818), SB\,815 (TIC\,169285097) have been first identified by \citet{1971AJ.....76..338S} 
as early type stars near the Southern Galactic pole and classified as sdB stars by \citet{1973AJ.....78..295G}. Both stars have been studied by the Montreal-Cambridge-Tololo survey 
as MCT\,0106$-$3259 and MCT\,2341$-$3443 \citep{2000AJ....119..241L}. PG\,0342+026 (TIC\,457168745) was discovered by the Palomar Green survey to be a sdB star \citep{1986ApJS...61..305G}. 
\citet{ostensen10c} were first attempting at finding pulsations in SB\,815. Unluckily, due to a short run, no variability has been reported. Another attempt was done using the SuperWASP telescope and \citep{holdsworth17} detected pulsations, marking the star as an sdBV.

We use Fourier technique to detect frequencies and asymptotic period spacings to identify pulsation modes and follows the first paper in our series \citep{charpinet19}. The work is a continuation of our effort started with the advent of the Kepler and K2 missions.

\section{Spectroscopic Analysis}
\begin{table*}
	\centering
	\caption{Basic information of the targets. First rows for each target refer to the parameters used in our work. Additional references are given for comparison. $^a$ Symmetric errors are given instead of the original asymmetric ones. Systematic errors are not accounted for.}
	\label{tab:table1}
	\begin{tabular}{ccccccccc} 
		\hline
		TIC & Name & Sectors & T$_{\rm eff}$[K] & $\log{g/(\rm cm\,s^{-2})}$ &$\log n_\text{He}/n_\text{H}$ & Gmag & distance [pc] & Reference for T$_{\rm eff}$ and $\log{g}$\\
		\hline
        067584818 & SB\,459 & 3 & 24,900(500) & 5.35(10) & -2.58(10) & 12.2 & 422(12) & \textbf{This work}\\
                  &         &   & 25,000(1200)& 5.30(20) & -2.8(30) &  &  &\citet{heber84}
                  \\
        169285097 & SB\,815 & 2 & 27,200(550)  & 5.39(10) & -2.94(01) & 10.9 & 246(5) & \citet{schneider18}\\
                  &         &   & 28,800(1500) & 5.40(20) & -2.46(30)& & & \citet{heber84}\\
                  &         &   & 28,390(300)$^a$  & 5.39(04)$^a$ & -3.07(24)$^a$& & &\citet{nemeth12}\\ 
                  &         &   & 27,000(1100) & 5.32(0.12)& -2.90(10)& &&\citet{geier13}\\
        457168745 & PG\,0342+026 & 5 & 26,000(1100) & 5.59(12) & -2.69(10) & 10.9 & 163(3)& \citet{geier13}
        \\
                  &              &   & 26200(1000) & 5.67(15)  & -2.4(15) & & &  \citet{saffer94}\\
        \hline
	\end{tabular}
\end{table*}
Atmospheric parameters 
of SB\,815 and PG\,0342+026 are available in the recent literature and are presented in Table\,\ref{tab:table1} along with our determination of these parameters of SB\,459. We also determined the radial velocities of PG\,0342+026, which is explained in details below.

In case of SB\,459, the only available quantitative spectral analysis was carried out by \citet{heber84}. 
Therefore, it was considered worthwhile to revisit the star and to take another spectrum with more advanced instruments than before. The ESO Faint Object Spectrograph and Camera\,2 (EFOSC2) spectrograph at the 3.58-metre New Technology Telescope (NTT) at the La Silla Observatory was used. The single spectrum was obtained on June 2019 with grism\,$\#$7, a slit of 1" covering the wavelength range from 3270\,\AA \,to \,5240\,\AA. Given that we used 2\,$\times$\,2 binning 
the nominal resolution of the spectrum should be $\Delta\lambda \sim 6.4$\,\AA. However, the seeing was excellent such that the slit was underfilled, which resulted in somewhat better resolution of 5.4\,\AA\ as measured directly from the spectrum.
The exposure time was 350\,seconds. We reduced the long-slit spectra using standard {\sc iraf} packages \citep{iraf1986,iraf1993}, by performing bias-subtraction, flat-field correction, wavelength and flux calibrations \citep{iraf1992,iraf1997}. The observed standard star was Feige\,110. The final spectrum has a signal to noise ratio of $\sim$\,300 at 4200 \AA.

SB\,459 was also observed with Boller \& Chivens Spectrograph at the 2.5-meter Ir\'en\'e Du Pont Telescope at the Las Campanas Observatory. 
The single spectrum was taken on 31 October 2019 using the following instrument setup, the grating of 600 lines/mm  corresponding to the central wavelength of 5000\,\AA\ covering a wider wavelength range from 3427 to 6573\,\AA. We used a slit width of 1" which resulted in somewhat better resolution, than EFOSC2, of $\Delta\lambda \sim 3.1$\,\AA. For the data reduction, we followed the same steps as in the case EFOSC2 spectra. The signal to noise ratio of the final spectrum is $\sim$\,250 at 4200 \AA\ with 600s exposure time.

\begin{figure}
\centering
\includegraphics[width=\columnwidth,angle=0]{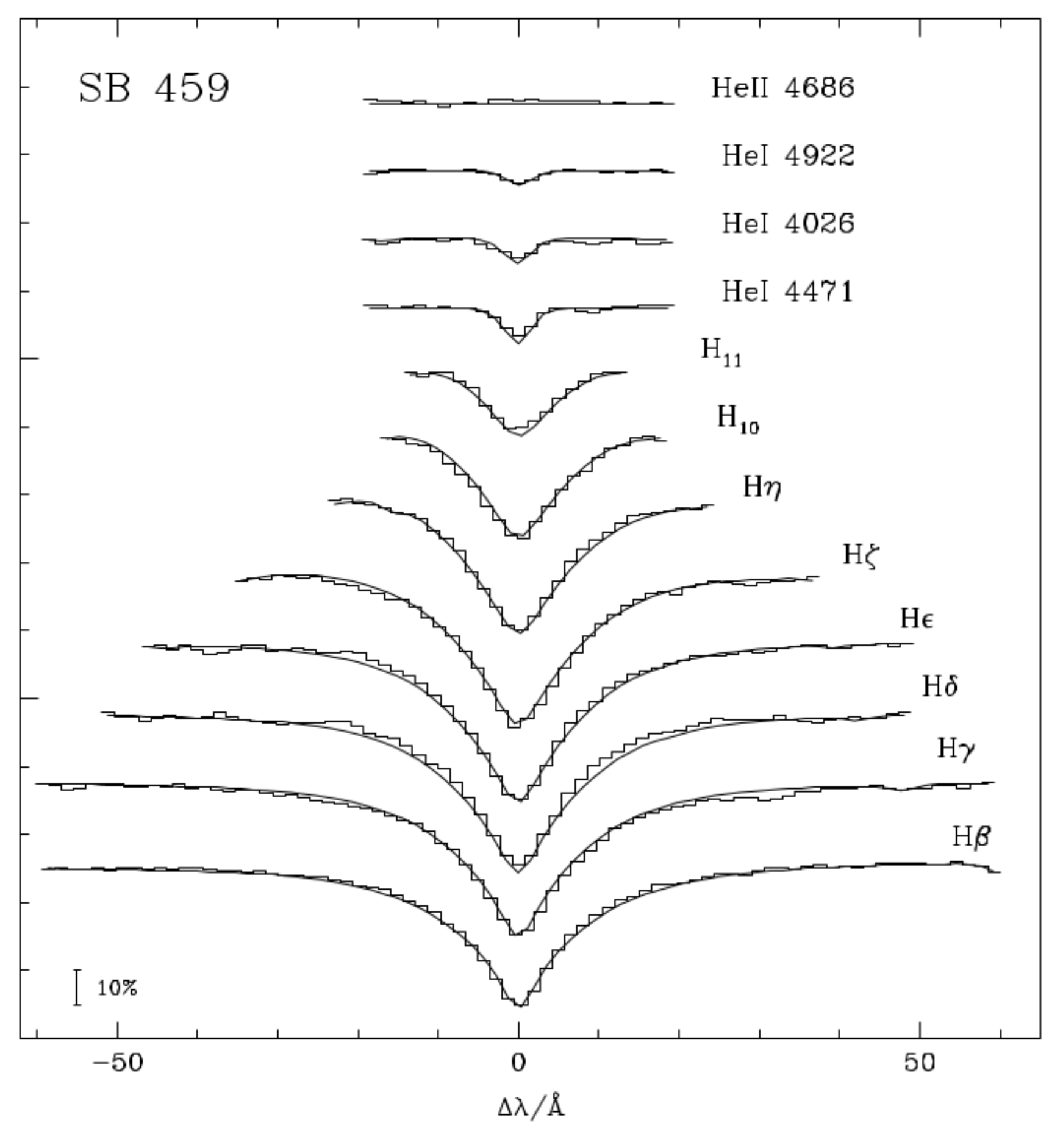}
\vspace{-4mm}
\caption{Spectral line fit of the Dupont spectrum of SB\,459.}
\label{fig_efosc_fit}
\end{figure}

We matched eight Balmer lines and four He\,I lines to both the EFOSC and the Dupont spectrum (Fig.\,\ref{fig_efosc_fit}) with the metal-line blanketed LTE grid of \citet{heber00} 
using $\chi^2$ minimisation techniques as described in  \citet{1999ApJ...517..399N}. 
The error budget is dominated by systematic errors, which we estimate at 2\% for the effective temperature and $\pm$0.1 dex for the surface gravity \citep[see][]{schneider18}. The resulting atmospheric parameters are remarkably similar at 
T$_{\rm eff}$\,=\,25,100\,K, $\log{g/({\rm cm\,s}^{-2})}$\,=\,5.34, $\log{n_{\rm He}/n_{\rm H}}$\,=\,-2.61 for the EFOSC spectrum and T$_{\rm eff}$\,=\,24,700\,K, $\log{g/({\rm cm}\,{\rm s}^{-2})}$\,=\,5.36, $\log{n_{\rm He}/n_{\rm H}}$\,=\,-2.55 for the Dupont spectrum. We adopted the mean values as listed in Table\,\ref{tab:table1}, which agree with the published values to within respective error limits.


\begin{figure}
\centering
\includegraphics[width=8.45cm,angle=0]{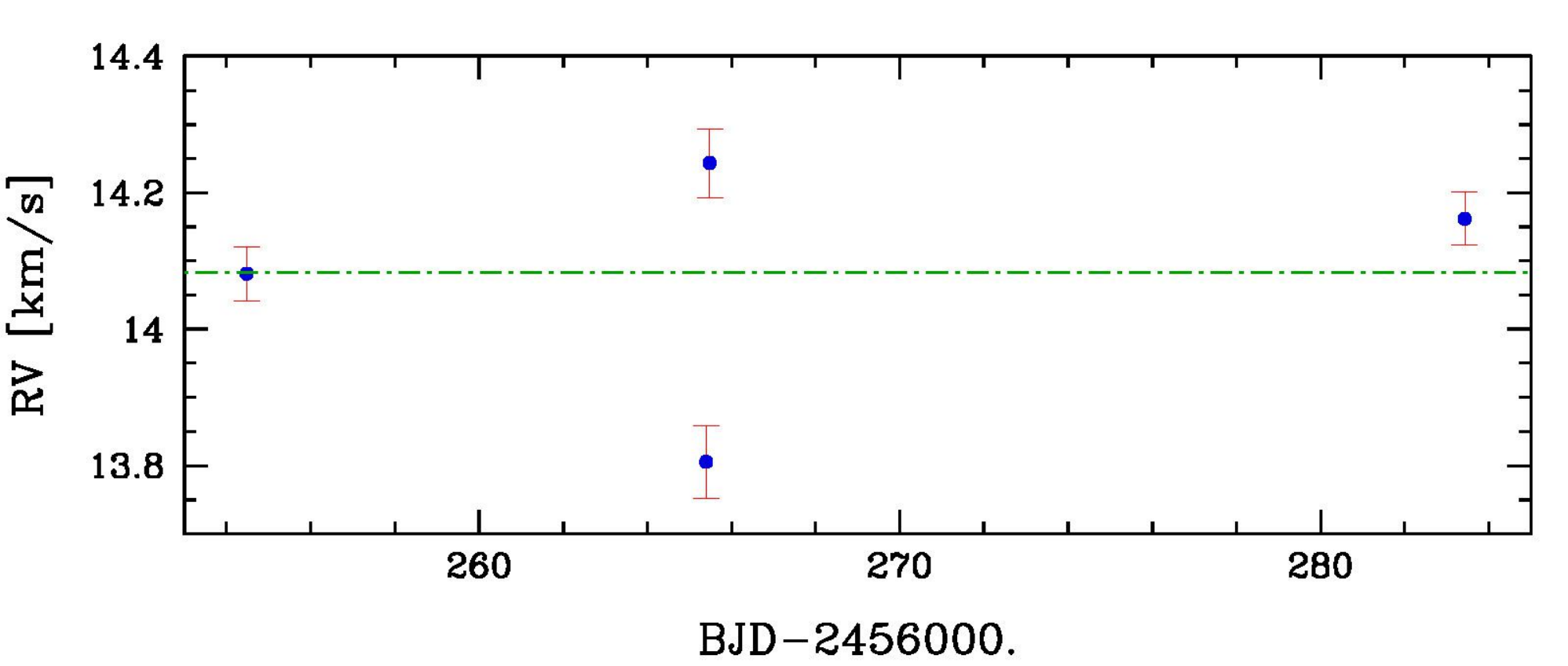}
\vspace{-4mm}
\caption{Radial velocities of PG\,0342+026. Although four measurements are not enough to fit the RV data with the main pulsation frequencies, at least they give a rough estimate of the RV amplitudes involved.}
\label{fig_rvs}
\end{figure}

\begin{figure}
\centering
\includegraphics[width=8.6cm,angle=0]{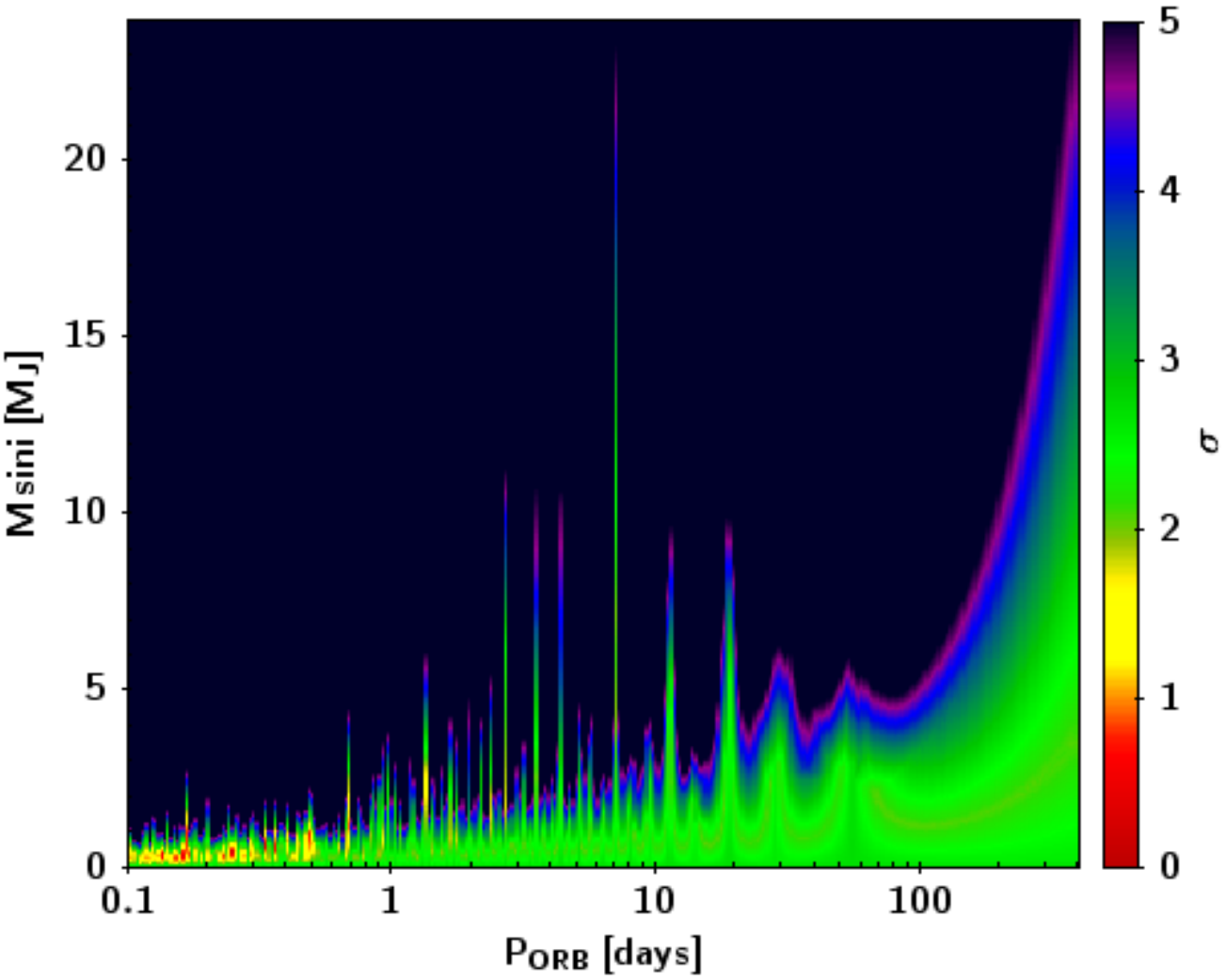}
\vspace{-4mm}
\caption{Upper limits to the mass of a hypothetical companion to PG\,0342+026 as a function of orbital period. The regions where the presence of a companion is more compatible with the RV measurements are those in red/yellow/green, while the regions in dark blue correspond to a very low probability to have a companion. See text for more details.}
\label{fig_comp}
\end{figure}

PG\,0342+026 was observed in November-December 2012 with Harps-N at the Telescopio Nazionale Galileo (TNG, La Palma) in the context of a program to search for sdB low-mass companions (see \citet{silvotti2020} for more details). Four high-resolution spectra were collected with a mean signal-to-noise ratio of 71 at 4700 \AA.\footnote{These spectra show a large number of metal absorption lines, many of which do not yet have a certain identification. The line identification is in progress and the results will be presented in a subsequent article.}
Using the cross correlation function on about 150 absorption lines (excluding H and He lines that are too broad), we computed the radial velocities (RV) of the star and we found a mean system velocity of +14.07\,km/s with significant variations around this value (Fig.\,\ref{fig_rvs}). Thanks to the TESS observations, we can now confirm that these variations are at least partly caused by g-mode pulsations, as it has been suspected since 2012. Having available only four RV data points, and knowing that this star pulsates in at least 20 frequencies, we are unable to obtain a reliable fit, however these data can be used to derive an upper limit to the minimum mass (M\,sin$i$) of a hypothetical companion. The question whether this sdB star is single or not is important for its evolution prior to EHB.

In order to set upper limits to the mass of a companion, we computed a series of synthetic RV curves for different orbital periods and companion masses, assuming circular orbits, and compared these curves with the RV measurements. For each synthetic RV curve we selected the phase that gives the best fit to the data using a weighted least squares algorithm. For each observational point we computed the difference, in absolute value and in $\upsigma$ units (where $\upsigma$ is the observation error), between observed and synthetic RV values. The color coding in Fig.\,\ref{fig_comp} corresponds to the mean value of this difference in $\upsigma$ units. We should keep in mind, however, that these upper limits to the mass of a companion are likely overestimated given that most if not all the variations that we see in Fig.\,\ref{fig_rvs} are likely caused by pulsations.


\section{Spectral energy distribution, interstellar reddening and stellar parameters}

Photometric measurements allow the angular diameters  to be determined along with the interstellar extinction, once the atmospheric parameters are known. We constructed spectral energy distributions from photometric measurements ranging from the ultraviolet (IUE) to the infrared. Infrared data were taken from 2MASS, VISTA-VIKING \citep[J,H,K; ][]{2006AJ....131.1163S} and WISE \citep[W1,W2, ][]{Cutri2012_WISE} catalogs. Magnitudes and colours in the Johnson \citep{1994AJ....107.1565A,1997A&AS..124..349M,2007AJ....133.2502L}, Str\"omgren 
\citep{1992AJ....104..203W,2015A&A...580A..23P}, 
APASS \citep{2015AAS...22533616H}, SkyMapper \citep{2018PASA...35...10W}, and Gaia \citep{2018A&A...616A...1G} photometric systems were fitted \citep[for details see][]{2018OAst...27...35H}. 
Three numerical box filters were defined to derive UV-magnitudes from IUE UV spectra covering the spectral ranges $1300$--$1800$\,\AA, $2000$--$2500$\,\AA, and $2500$--$3000$\,\AA. Interstellar extinction is accounted for using the extinction curve of  \citet{1999PASP..111...63F}. 

\begin{figure}
\begin{center}
\includegraphics[width=\columnwidth]{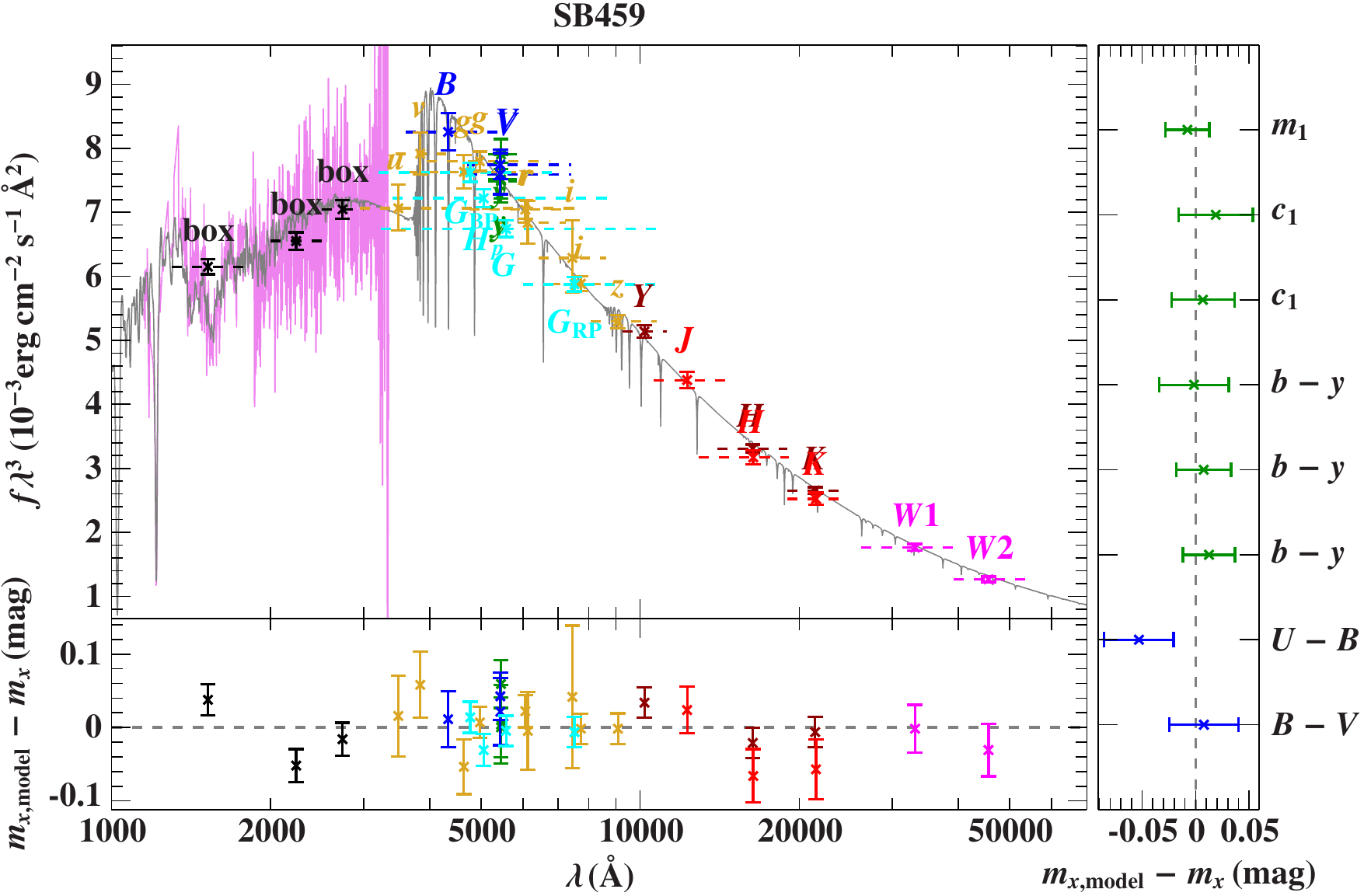}
\includegraphics[width=\columnwidth]{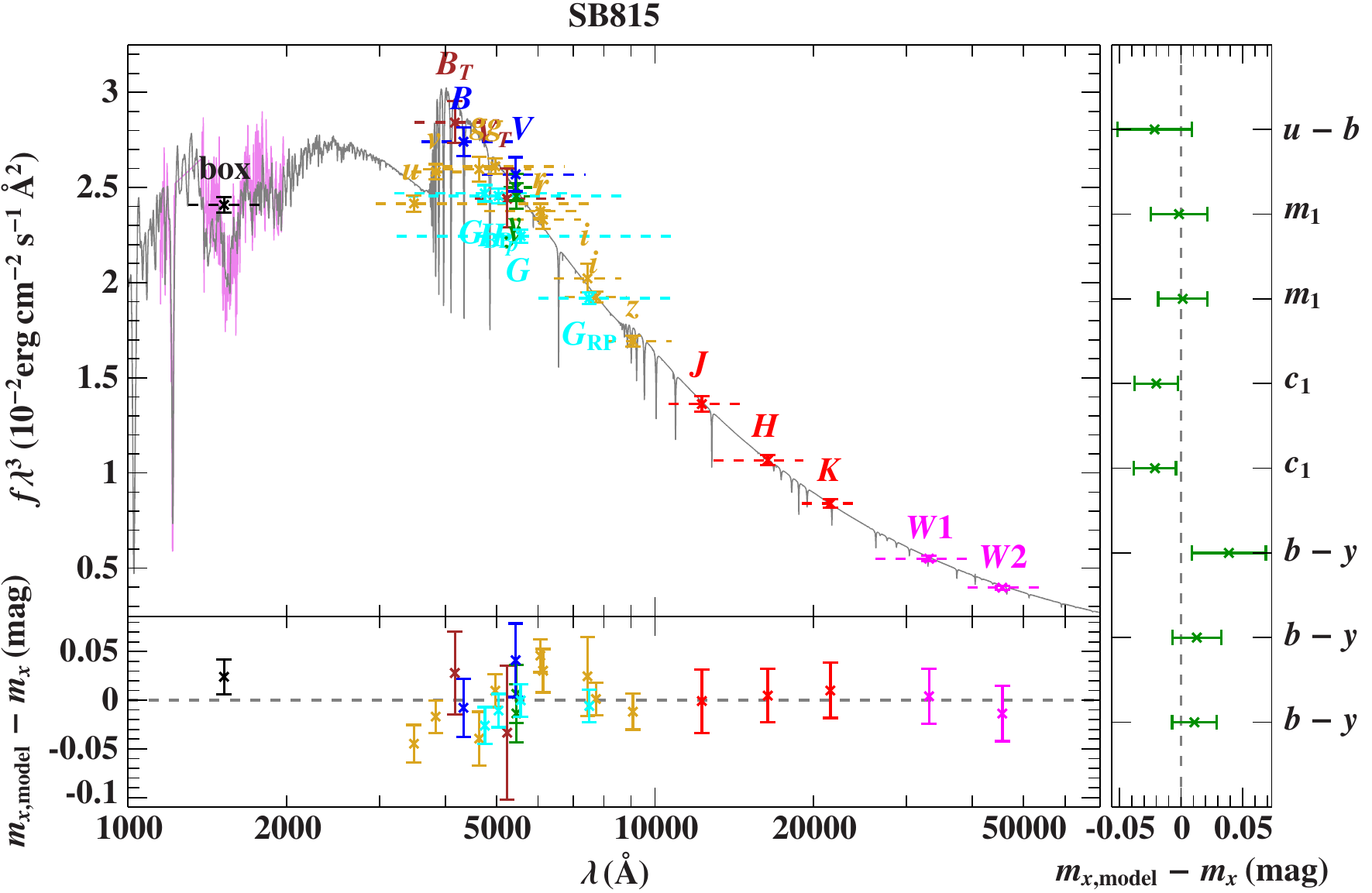}
\includegraphics[width=\columnwidth]{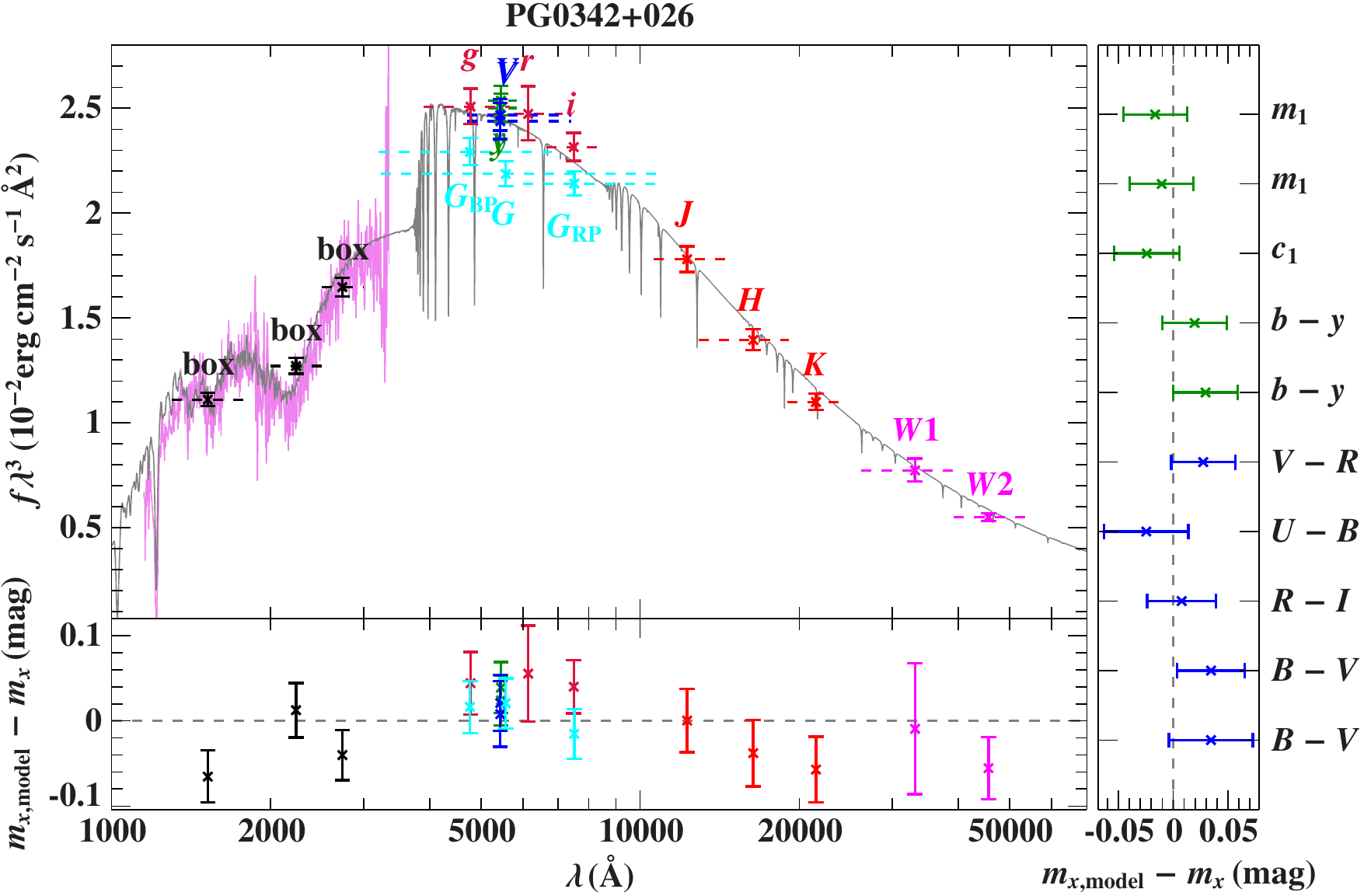}
\caption{Matching the spectral energy distributions and colours of SB\,459 (upper panel), SB\,815 (middle panel), and PG0342+026\ (lower panel). The colored observed magnitudes (IUE box: black, SDSS: blue; SkyMapper: yellow; {\it Gaia}: cyan; VISTA: dark red; 2MASS: red, WISE:magenta) were derived from filter-averaged fluxes. The dashed horizontal lines indicate the filter widths). A model SED calculated with the spectroscopic parameters is overplotted as a solid gray line. Also overlayed are the IUE spectra in magenta. The panels below (SED) and to the right (colours) of the main panel show the residuals for the magnitudes and colours.}
\label{fig_sed_fits}
\end{center}
\end{figure}

The angular diameter and the interstellar reddening parameter E(B-V) were the only free parameter in the matching of the synthetic the SEDs to the observed ones. In Figure \ref{fig_sed_fits} we plot the SEDs as flux density times the wavelength to the power of three (F$_\lambda \lambda^3$) versus the wavelength to reduce the steep slope of the SED over such a broad wavelength range. We also display the residuals (O-C) of the magnitudes and the colours. The synthetic SEDs match the observed ones very well in all parts of the wide spectral range. Hence, there is no contribution from potential companions at any wavelength for all three stars. Interstellar reddening is consistent with zero for SB\,459 and SB\,815 and small for PG\,0342+026, all in accordance with the predictions of the maps of \citet{1998ApJ...500..525S} and \citet{2011ApJ...737..103S}. The resulting angular diameters and interstellar reddening parameters are given in Tables \ref{tab:sed_results_sb459}, \ref{tab:sed_results_sb815}, and \ref{tab:sed_results_pg0342+026}.

\begin{table}
	\centering
	\caption{SB\,459: Angular diameters, interstellar reddening parameter, Gaia parallax, and stellar parameters.
	}
	\label{tab:sed_results_sb459}
	\begin{tabular}{lr}
\hline\hline
Object: SB459 & 68\% confidence interval\\
\hline
Angular diameter $\log(\Theta\,\mathrm{(rad)})$ & $-10.6134\pm0.0016$ \\
Color excess $E(B-V)$ & $0.0049\pm0.0021$\,mag \\
Parallax $\varpi$ ({\it Gaia}, $\text{RUWE}=1.17$) & $2.37\pm0.07$\,mas \\
Effective temperature $T_{\mathrm{eff}}$ (prescribed) & $24900\pm500$\,K \\
Surface gravity $\log (g\,\mathrm{(cm\,s^{-2})})$ (prescribed) & $5.35\pm0.10$ \\
Helium abundance $\log(n(\textnormal{He}))$ (fixed) & $-2.58$ \\
Radius $R_\star$ & $0.228\pm0.007$\,$R_\odot$ \\
Mass $M_\star$& $0.42\pm0.11$\,$M_\odot$ \\
Luminosity $\log\left(\frac{L}{L_\odot}\right)$ & $1.25\pm0.05$ \\
\hline
\end{tabular}

\end{table}

\begin{table}
	\centering
	\caption{Same as Table \ref{tab:sed_results_sb459}, but for SB\,815
	}
	\label{tab:sed_results_sb815}
	\begin{tabular}{lr}
\hline\hline
Object: SB815 & 68\% confidence interval\\
\hline
Angular diameter $\log(\Theta\,\mathrm{(rad)})$ & $-10.3920^{+0.0018}_{-0.0017}$ \\
Color excess $E(B-V)$ & $0.0018^{+0.0025}_{-0.0018}$\,mag \\
Parallax $\varpi$ ({\it Gaia}, $\text{RUWE}=1.23$) & $4.07\pm0.10$\,mas \\
Effective temperature $T_{\mathrm{eff}}$ (prescribed) & $27200\pm550$\,K \\
Surface gravity $\log (g\,\mathrm{(cm\,s^{-2})})$ (prescribed) & $5.39\pm0.10$ \\
Helium abundance $\log(n(\textnormal{He}))$ (fixed) & $-2.94$ \\
Radius $R_\star$ & $0.221\pm0.005$\,$R_\odot$ \\
Mass $M_\star$ & $0.44\pm0.11$\,$M_\odot$ \\
Luminosity $\log\left(\frac{L}{L_\odot}\right)$ & $1.38\pm0.05$ \\
\hline
\end{tabular}

\end{table}

\begin{table}
	\centering
	\caption{Same as Table \ref{tab:sed_results_sb459}, but for PG0342+026}
	\label{tab:sed_results_pg0342+026}
	\begin{tabular}{lr}
\hline\hline
Object: PG0342+026 & 68\% confidence interval\\
\hline
Angular diameter $\log(\Theta\,\mathrm{(rad)})$ & $-10.2975\pm0.0025$ \\
Color excess $E(B-V)$ & $0.128\pm0.004$\,mag \\
Parallax $\varpi$ ({\it Gaia}, $\text{RUWE}=1.28$) & $6.13\pm0.13$\,mas \\
Effective temperature $T_{\mathrm{eff}}$ (prescribed) & $26000\pm1100$\,K \\
Surface gravity $\log (g\,\mathrm{(cm\,s^{-2})})$ (prescribed) & $5.59\pm0.12$ \\
Helium abundance $\log(n(\textnormal{He}))$ (fixed) & $-2.69$ \\
Radius $R_\star$ 
& $0.182\pm0.004$\,$R_\odot$ \\
Mass $M_\star$
& $0.47\pm0.14$\,$M_\odot$ \\
Luminosity $\log\left(\frac{L}{L_\odot}\right)$ 
& $1.13\pm0.08$ \\
\hline
\end{tabular}

\end{table}


In its second data release the Gaia mission \citep{2018A&A...616A...1G} provided trigonometric parallaxes of high precision (to better than 3\%) for all three stars. The ``renormalized unit weight error'' \citep[RUWE, see][]{RUWE} is a good quality indicator for the astrometric solution, because it is independent of the color of the object. This makes it the best choice to judge the quality of the Gaia parallaxes of blue stars, such as studied here. The RUWE value is below the recommended value of $1.4$ for all three stars,
indicating that the astrometric solutions are reliable. The Gaia parallaxes and the angular diameters allow us to convert the atmospheric parameters to stellar radii via $R_\star\,=\,\Theta/(2\varpi)$, masses via $M_\star=g R_\star^2/G$, and luminosities via $\log\left(\frac{L}{L_\odot}\right)\,=\,\log\left(\left(\frac{R_\star}{R_\odot}\right)^2\left(\frac{T_{\mathrm{eff}}}{5775\,\mathrm{K}}\right)^4\right)$. The results are summarized in Tables\,\ref{tab:sed_results_sb459}, \ref{tab:sed_results_sb815}, and \ref{tab:sed_results_pg0342+026}. Uncertainties of the derived radii and luminosities are small because of the high precision of the Gaia parallaxes and well constrained effective temperatures. The derived masses, however, have larger uncertainties resulting from the uncertainties of the spectroscopic surface gravities. The resulting masses are close to canonical \citep{1993ApJ...419..596D}, but uncertainties are large, mainly due to the surface gravity not yet being sufficiently constrained.

\section{Light variations}
\begin{figure*}
	\includegraphics[width=\textwidth]{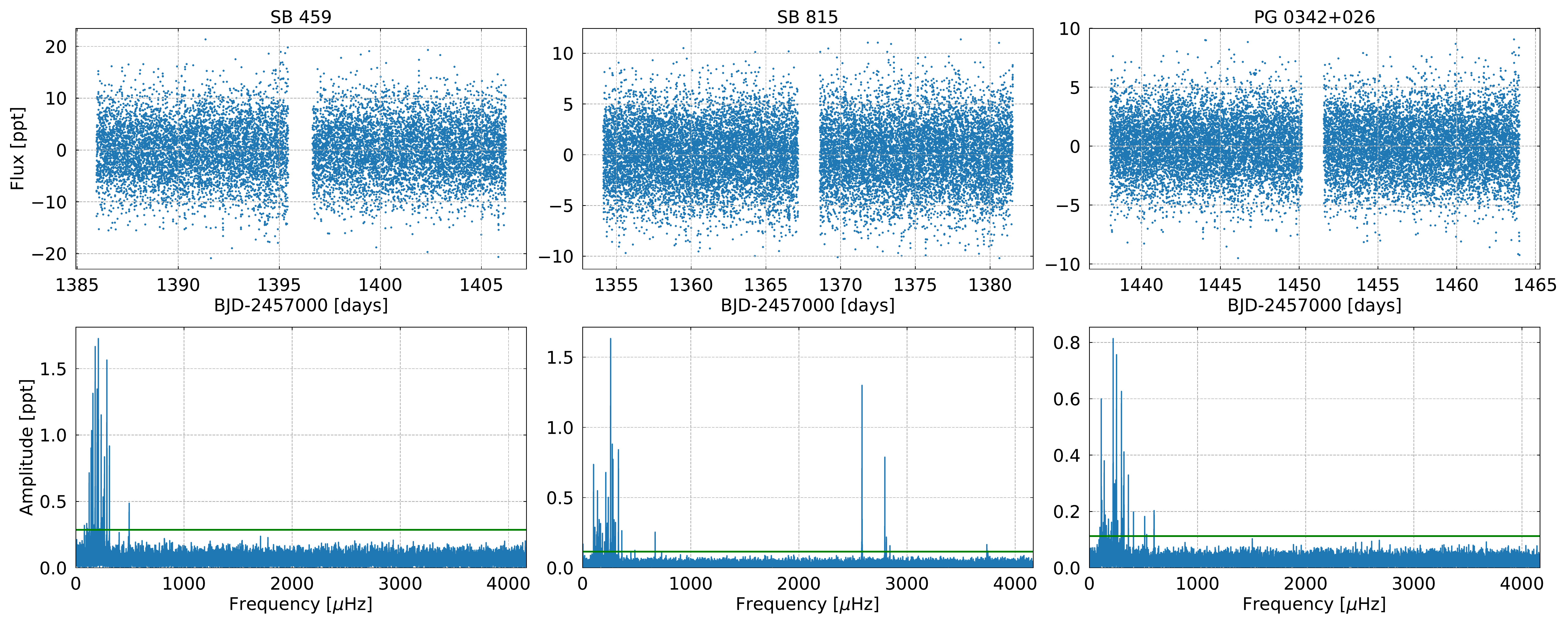}
    \caption{The upper panels show the light curves, while the bottom panels show the amplitude spectra. The green horizontal line in the bottom panels denotes 4.5$\upsigma$ threshold.}
    \label{fig:figure1}
\end{figure*}

All three targets have been observed during single TESS sectors, which are specifically listed in Table\,\ref{tab:table1},
and have been observed in the short cadence (SC) mode, lasting 120\,s. We performed our analysis by using the corrected time series data extracted through the TESS data processing pipeline developed by NASA's Science Processing Operation Centre. These processed data are publicly available in the Mikulski Archive for Space Telescopes database. We collected these files and have done further analysis. We extracted PDCSAP\_FLUX, which is corrected for on-board systematics and neighbors' contribution to the overall flux. We clipped fluxes at 5$\upsigma$ to remove outliers, de-trended long term variation (longer than days). Finally, we normalized fluxes by calculating $(f/<f>-1)*1000$, deriving {\it part per thousand} (ppt). We show the resultant light curves of each target in the top panels of Figure\,\ref{fig:figure1}.

\subsection{Fourier Analysis}
We used a Fourier technique for identifying the frequency of pulsations. Since we have only about 27\,days long SC data for each target, the frequency resolution is 0.64\,${\mu}$Hz as defined by 1.5/T, where T is the time coverage of the data \citep{baran12c}. The standard prewhitening procedure has been used by fitting the peaks with $A_i\sin(2{\pi}f_it+{\phi_i})$ by means of a non-linear least-square method. We have used our custom pipeline for this purpose. We prewhitened the data down to the detection threshold defined as 4.5 times the mean noise level, i.e. the signal to noise ratio, S/N\,=\,4.5, calculated from the residual amplitude spectra. The threshold has been discussed by \citet{baran15,zong16}, who reported slightly higher threshold than in case of ground-based data.
The SC mode sampling translates to the Nyquist frequency of 4166\,$\upmu$Hz. In case of SB\,815 we found some frequencies in the p-mode range, close to the Nyquist frequency and, following a discovery of a reflection across Nyquist \citep{baran12a}, we also searched the amplitude spectrum above the Nyquist frequency in order to see if any of subNyquist p-modes have superNyquist origin. An amplitude and a profile of the peaks in the sub and superNyquist regions help identifying the origin of the signal. Since the Nyquist frequency is not fixed in time, the reflections will look smeared out and therefore lower in amplitude, however this effect works best if the satellite motion covers substantial part of the orbit, which does not happen in case of single sector data. Therefore, we assume that all frequencies in the p-mode region originate in the subNyquist range are real, but this should be confirmed with shorter cadence data, presumably during the second phase of TESS mission, when 20\,sec cadence will be accessible.

In SB\,459, we detected 22 frequencies above the detection threshold with 207.314\,$\upmu$Hz being the highest amplitude one at 1.72\,ppt, and all are in the g-mode region. In SB\,815 we detected 37 frequencies in the g-mode region and six frequencies in the p-mode region, with the highest amplitude (1.642\,ppt) frequency at 258.1878\,$\upmu$Hz. In PG\,0342+026, we detected 27 frequencies, with 219.274\,$\upmu$Hz having the highest amplitude of 0.758\,ppt. We list all frequencies detected in those three stars in Tables\,\ref{tab:table2}, \ref{tab:table3} and \ref{tab:table4}. 

SB\,815 turned out to be a hybrid pulsator. The highest amplitude of 1.296\,ppt in the p-mode region shows at 2582.8740\,$\upmu$Hz. The signal at high frequencies is parted into two groups. The first one contains four frequencies, while the second one has two low amplitude frequencies. We found the separation between these two groups to be around 896\,$\upmu$Hz. Such spacing has been previously reported by \citep{baran09}, who concluded that such groups may represent modes with two consecutive radial orders.

\begin{table}
\centering
\caption{List of frequencies detected in SB\,459}
\label{tab:table2}
\begin{tabular}{ccccccc}
\hline
\multirow{2}{*}{ID} & Frequency & Period & Amplitude & \multirow{2}{*}{S/N} & \multirow{2}{*}{\it l} & \multirow{2}{*}{n}\\
&[$\upmu$Hz] & [s] & [ppt] &&\\
\hline
f$_{\rm 1}$ &   77.82(5) &  12849.8(9.0) &   0.30(5) &   4.5 & &\\
f$_{\rm 2}$ &  100.092(48) &   9990.8(4.7) &   0.34(5) &   5.2 & 1 & 39\\
f$_{\rm 3}$ &  122.319(22) &   8175.3(1.5) &   0.74(5) &  11.1 & 1/2 & 32/55\\
f$_{\rm 4}$ &  126.27(5) &   7919.5(3.3) &   0.31(5) &   4.6 & 1 & 31\\
f$_{\rm 5}$ &  140.180(19) &   7133.7(10) &   0.86(5) &  12.9 & 1 & 28\\
f$_{\rm 6}$ &  145.213(17) &   6886.4(8) &   0.97(5) &  14.6 & 1 & 27\\
f$_{\rm 7}$ &  156.802(13) &   6377.5(5) &   1.30(5) &  19.6 & 1/2 & 25/43\\
f$_{\rm 8}$ &  163.519(48) &   6115.5(1.8) &   0.34(5) &   5.1 & 1 & 24\\
f$_{\rm 9}$ &  178.170(10) &   5612.63(31) &   1.66(5) &  24.9 & 1 & 22\\
f$_{\rm 10}$ &  196.322(12) &   5093.67(31) &   1.36(5) &  20.4 & 1 & 20\\
f$_{\rm 11}$ &  207.314(9) &   4823.61(22) &   1.72(5) &  25.8 & 1 & 19\\
f$_{\rm 12}$ &  221.87(5) &   4507.2(1.1) &   0.30(5) &   4.6 & &\\
f$_{\rm 13}$ &  233.175(14) &   4288.63(26) &   1.17(5) &  17.6 & 1 & 17\\
f$_{\rm 14}$ &  234.566(42) &   4263.2(8) &   0.39(5) &   5.9 & &\\
f$_{\rm 15}$ &  242.460(42) &   4124.4(7) &   0.39(5) &   5.9 & 2 & 28\\
f$_{\rm 16}$ &  246.917(46) &   4049.9(8) &   0.35(5) &   5.3 & 1 & 16\\
f$_{\rm 17}$ &  251.341(32) &   3978.7(5) &   0.51(5) &   7.7 & 2 & 27\\
f$_{\rm 18}$ &  261.184(28) &   3828.72(40) &   0.60(5) &   9.0 & 2 & 26\\
f$_{\rm 19}$ &  263.905(20) &   3789.24(28) &   0.84(5) &  12.6 & 1 & 15\\
f$_{\rm 20}$ &  286.201(10) &   3494.05(13) &   1.58(5) &  23.7 & 1 & 14\\
f$_{\rm 21}$ &  309.779(17) &   3228.11(18) &   0.93(5) &  14.0 & 1/2 & 13/22\\
f$_{\rm 22}$ &  492.395(34) &   2030.89(14) &   0.47(5) &   7.1 & 2 & 14\\
\hline
\end{tabular}
\end{table}

\begin{table}
	\centering
	\caption{List of frequencies detected in SB\,815}
	\label{tab:table3}
\begin{tabular}{ccccccc}
\hline
\multirow{2}{*}{ID} & Frequency & Period & Amplitude & \multirow{2}{*}{S/N} & \multirow{2}{*}{\it l} & \multirow{2}{*}{n}\\
&[$\upmu$Hz] & [s] & [ppt] &&\\
\hline
f$_{\rm 1}$ &  100.438(7) &   9956.4(7) &   0.718(21) &  27.3 & 1 & 38\\
f$_{\rm 2}$ &  103.574(39) &   9655.0(3.6) &   0.125(21) &   4.8 & 1 & 37\\
f$_{\rm 3}$ &  106.159(34) &   9419.8(3.0) &   0.141(21) &   5.4 & 1/2 & 36/62\\
f$_{\rm 4}$ &  112.435(31) &   8894.0(2.4) &   0.228(22) &   8.7 & &\\
f$_{\rm 5}$ &  112.789(24) &   8866.1(1.9) &   0.291(22) &  11.1 & 1 & 34\\
f$_{\rm 6}$ &  123.734(32) &   8081.9(2.1) &   0.149(21) &   5.7 & 1 & 31\\
f$_{\rm 7}$ &  128.523(22) &   7780.7(1.3) &   0.217(21) &   8.3 & 1 & 30/t\\
f$_{\rm 8}$ &  131.737(20) &   7590.9(1.2) &   0.242(21) &   9.2 & 1/2 & 29/50\\
f$_{\rm 9}$ &  136.885(9) &   7305.4(5) &   0.534(22) &  20.3 & 1 & 28\\
f$_{\rm 10}$ &  137.674(27) &   7263.5(1.4) &   0.186(22) &   7.1 & 2 & 48\\
f$_{\rm 11}$ &  142.334(26) &   7025.7(1.3) &   0.193(22) &   7.3 & 1 & 27\\
f$_{\rm 12}$ &  142.858(26) &   6999.9(1.3) &   0.189(22) &   7.2 & &\\
f$_{\rm 13}$ &  151.999(14) &   6579.0(6) &   0.345(21) &  13.1 & 1 & 25/t\\
f$_{\rm 14}$ &  154.178(37) &   6486.0(1.5) &   0.132(21) &   5.0 & &\\
f$_{\rm 15}$ &  165.197(16) &   6053.4(6) &   0.306(21) &  11.6 & 2 & 40\\
f$_{\rm 16}$ &  174.841(36) &   5719.5(1.2) &   0.133(21) &   5.1 & 1 & 22\\
f$_{\rm 17}$ &  182.576(20) &   5477.2(6) &   0.238(21) &   9.0 & 1 & 21\\
f$_{\rm 18}$ &  202.345(27) &   4942.1(7) &   0.179(21) &   6.8 & 1 & 19\\
f$_{\rm 19}$ &  213.908(7) &   4674.92(16) &   0.671(21) &  25.6 & 1/2 & 18/31\\
f$_{\rm 20}$ &  226.812(15) &   4408.93(29) &   0.328(21) &  12.5 & 1 & 17\\
f$_{\rm 21}$ &  228.836(39) &   4369.9(7) &   0.123(21) &   4.7 & 2 & 29\\
f$_{\rm 22}$ &  236.890(10) &   4221.36(17) &   0.489(21) &  18.6 & 2 & 28\\
f$_{\rm 23}$ &  246.268(38) &   4060.6(6) &   0.125(21) &   4.8 & 2 & 27\\
f$_{\rm 24}$ &  258.1879(29) &   3873.149(44) &   1.642(21) &  62.5 & 1 & 15\\
f$_{\rm 25}$ &  266.359(24) &   3754.33(33) &   0.204(21) &   7.8 & 2 & 25\\
f$_{\rm 26}$ &  273.537(5) &   3655.82(7) &   0.878(21) &  33.4 & 1 & t\\
f$_{\rm 27}$ &  277.625(34) &   3601.99(44) &   0.143(21) &   5.4 & 2 & 24\\
f$_{\rm 28}$ &  279.723(6) &   3574.96(8) &   0.777(21) &  29.6 & 1 & 14\\
f$_{\rm 29}$ &  285.303(34) &   3505.05(42) &   0.141(21) &   5.4 & &\\
f$_{\rm 30}$ &  289.809(14) &   3450.55(16) &   0.353(21) &  13.4 & 2 & 23\\
f$_{\rm 31}$ &  302.183(37) &   3309.26(40) &   0.137(22) &   5.2 & &\\
f$_{\rm 32}$ &  302.892(14) &   3301.51(16) &   0.353(22) &  13.4 & 1/2 & 13/22\\
f$_{\rm 33}$ &  330.565(6) &   3025.12(5) &   0.848(21) &  32.3 & 1 & 12\\
f$_{\rm 34}$ &  361.604(19) &   2765.46(14) &   0.257(21) &   9.8 & 1 & 11\\
f$_{\rm 35}$ &  445.775(40) &   2243.29(20) &   0.121(21) &   4.6 & 1 & 9\\
f$_{\rm 36}$ &  482.523(40) &   2072.44(17) &   0.119(21) &   4.5 & 2 & 14\\
f$_{\rm 37}$ &  669.836(19) &   1492.902(42) &   0.255(21) &   9.7 & &\\
f$_{\rm 38}$ & 2582.8740(37) &    387.1656(6) &   1.296(21) &  49.3 & &\\
f$_{\rm 39}$ & 2793.905(6) &    357.9219(8) &   0.786(21) &  29.9 & &\\
f$_{\rm 40}$ & 2808.165(22) &    356.1045(28) &   0.214(21) &   8.1 & &\\
f$_{\rm 41}$ & 2841.082(31) &    351.9786(39) &   0.153(21) &   5.8 & &\\
f$_{\rm 42}$ & 3737.134(28) &    267.5848(20) &   0.169(21) &   6.4 & &\\
f$_{\rm 43}$ & 3747.579(40) &    266.8390(29) &   0.118(21) &   4.5 & &\\
\hline
\end{tabular}
\end{table}

\begin{table}
\centering
\caption{List of frequencies detected in PG\,0342+026}
\label{tab:table4}
\begin{tabular}{ccccccc}
\hline
\multirow{2}{*}{ID} & Frequency & Period & Amplitude & \multirow{2}{*}{S/N} & \multirow{2}{*}{\it l} & \multirow{2}{*}{n}\\
&[$\upmu$Hz] & [s] & [ppt] &&\\
\hline
f$_{\rm 1}$ &   96.786(35) &  10332.1(3.7) &   0.145(21) &   5.5 & &\\
f$_{\rm 2}$ &  108.889(9) &   9183.7(7) &   0.593(21) &  22.6 & 1/2 & 40/69\\
f$_{\rm 3}$ &  114.621(23) &   8724.4(1.7) &   0.222(21) &   8.5 & 1 & 38\\
f$_{\rm 4}$ &  124.720(42) &   8018.0(2.7) &   0.120(21) &   4.6 & 1 & 35\\
f$_{\rm 5}$ &  128.559(32) &   7778.5(1.9) &   0.161(21) &   6.1 & 1 & 34\\
f$_{\rm 6}$ &  132.313(32) &   7557.8(1.8) &   0.160(21) &   6.1 & 1 & 33\\
f$_{\rm 7}$ &  136.448(13) &   7328.8(7) &   0.391(21) &  14.9 & 1 & 32\\
f$_{\rm 8}$ &  145.763(28) &   6860.5(1.3) &   0.185(21) &   7.0 & 1 & 30\\
f$_{\rm 9}$ &  150.803(39) &   6631.2(1.7) &   0.131(21) &   5.0 & 1 & 29\\
f$_{\rm 10}$ &  156.352(34) &   6395.8(1.4) &   0.152(21) &   5.8 & 1 & 28\\
f$_{\rm 11}$ &  175.286(30) &   5705.0(10) &   0.170(21) &   6.5 & 1/2 & 25/43\\
f$_{\rm 12}$ &  198.546(36) &   5036.6(9) &   0.141(21) &   5.4 & 2 & 38\\
f$_{\rm 13}$ &  204.375(30) &   4893.0(7) &   0.167(21) &   6.3 & 2 & 37\\
f$_{\rm 14}$ &  219.274(6) &   4560.50(13) &   0.819(21) &  31.1 & 1 & 20\\
f$_{\rm 15}$ &  231.527(17) &   4319.16(32) &   0.298(21) &  11.3 & 1 & 19\\
f$_{\rm 16}$ &  243.921(17) &   4099.69(28) &   0.307(21) &  11.7 & 1/2 & 18/31\\
f$_{\rm 17}$ &  250.256(7) &   3995.90(11) &   0.758(21) &  28.8 & 1 & t\\
f$_{\rm 18}$ &  260.620(30) &   3837.00(45) &   0.167(21) &   6.4 & 1/2 & 17/29\\
f$_{\rm 19}$ &  295.891(8) &   3379.62(9) &   0.623(21) &  23.7 & 1 & 15\\
f$_{\rm 20}$ &  303.246(28) &   3297.65(30) &   0.181(21) &   6.9 & 2 & 25\\
f$_{\rm 21}$ &  315.536(20) &   3169.21(20) &   0.252(21) &   9.6 & 2 & 24\\
f$_{\rm 22}$ &  318.907(13) &   3135.71(13) &   0.390(21) &  14.8 & 1 & 14\\
f$_{\rm 23}$ &  359.953(15) &   2778.14(12) &   0.329(21) &  12.5 & 1 & t\\
f$_{\rm 24}$ &  406.920(26) &   2457.48(16) &   0.194(21) &   7.4 & 1 & 11\\
f$_{\rm 25}$ &  510.726(28) &   1958.00(11) &   0.182(21) &   6.9 & 1/2 & 9/15\\
f$_{\rm 26}$ &  529.408(43) &   1888.90(15) &   0.119(21) &   4.5 & &\\
f$_{\rm 27}$ &  597.478(25) &   1673.70(7) &   0.204(21) &   7.8 & &\\
\hline
\end{tabular}
\end{table}

\subsection{Multiplets}
Multiplets are a result of stellar rotation that changes frequency of modes with the same modal degree and {\it m}$\neq$0. The frequency change also depends on a rotation period of a star. For a given modal degree {\it l} there is 2{\it l}\,+\,1 components differing in an azimuthal order {\it m}, therefore by the number of components in an identified multiplet we can infer the modal degree.

We could not detect multiplets in any of these three targets. The reason for null detection may be not long enough data coverage, which causes the frequency resolution not to be high enough to resolve multiplet components. A common rotation period derived in sdB stars is around 40\,days \citep[e.g.][]{baran12a,baran12b,telting12,ostensen14,foster15,charpinet18}, which, in case of p-modes, translates to 0.29$\upmu$Hz or half the frequency resolution of our data, though exceptions are found \citep{baran09,reed14}. Another explanation may be a pole-on orientation of a pulsation axis, however we consider this explanation to be very unlikely, since we do not expect all three randomly chosen targets to be oriented in exactly the same way. In case the amplitudes of the side components are low, below the detection threshold, these components will not be detected, either.

\subsection{Asymptotic Period Spacing}
Another method that helps identifying modes relies on periods and not frequencies. In the asymptotic limit, {\it i.e. n$\gg$l}, consecutive overtones of g-modes are nearly equally spaced in period \citep[e.g.][]{charpinet00,reed11}. The pulsation period of a given mode with degree {\it l} and radial order {\it n} can be expressed as
\begin{equation}
    P_{l,n}=\frac{P_0}{\sqrt{l(l+1)}}n+\epsilon
	\label{eq:1}
\end{equation}
where $P_0$ is the period of the fundamental radial mode and $\epsilon$ is an offset \citep{unno79}. Thus, for two consecutive radial overtones and a given modal degree, a difference (commonly called as {\it period spacing}) of their periods should be constant, dependent of the modal degree and independent of the radial order.
\begin{equation}
    {\Delta}P_l=P_{l,n+1}-P_{l,n}=\frac{P_0}{\sqrt{l(l+1)}}
	\label{eq:2}
\end{equation}
Using Equation\,\ref{eq:1} it is possible to assign the radial order {\it n} to the precision of some arbitrarily chosen offset $n_l$. We provide those values in Tables\,\ref{tab:table2},\ref{tab:table3} and \ref{tab:table4}. Using Equation\,\ref{eq:2} we can also derive a ratio between a period spacing of modes of different modal degree, {\it e.g.} the ratio between dipole and quadrupole modes equals 1/$\sqrt{\rm 3}$. This is very strong constraint, since having the period spacing for dipole modes, we can estimate the expected value for higher degree modes. Previous analyses of photometric space data of sdBVs show that the average period spacing of dipole modes is nearly 250\,s on average \citep{reed18b}. The average spacing for quadrupole modes is found to be close to the expected value, being a result of the ratio given above.

\begin{figure*}
\includegraphics[width=\textwidth]{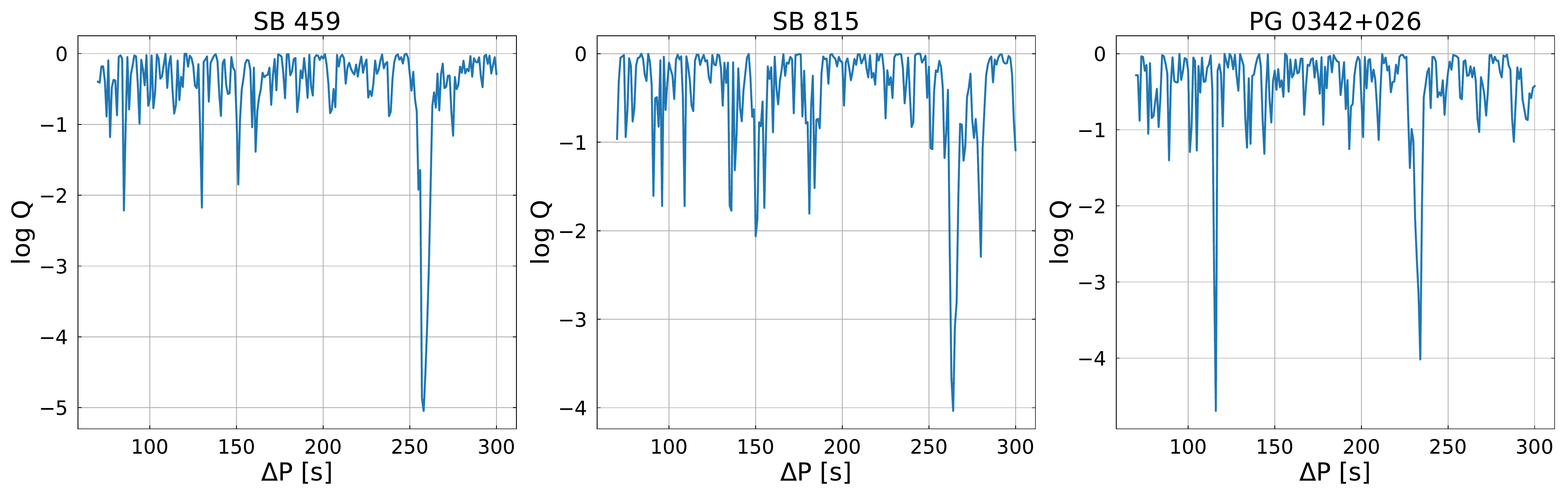}
\caption{Kolmogorov-Smirnov test for SB\,459, SB\,815 and PG\,0342+026}
\label{fig:kstest}
\end{figure*}

Best, if the mode identification is done based on both features, multiplets and period spacing, since they complement each other providing more convincing conclusion on a mode assignment, very often helping to start finding a specific modal degree sequence. In our case, we had to rely solely on the period spacing. We started our modal degree assignment with the highest amplitude modes. This assumption is justified by the surface cancellation effect, which causes that modes with higher degrees have smaller observational amplitudes. In this consideration, it is assumed that all modes have the same intrinsic amplitudes, which may not necessarily be correct, however our thus far experience clearly shows that most of the high amplitude frequencies in sdBV stars are dipole modes. Despite of this assumption, if two peaks satisfy both dipole and quadrupole sequences, we mentioned both values in tables and figures. In \'echelle and reduced period diagrams we have added these points with different color coding. The average period spacing in sdBVs detected thus far is between 200 and 300\,sec. To guess the average spacing in our targets, we calculated the {\it Kolmogorov-Smirnov} (KS) test and we plotted the results in Figure\,\ref{fig:kstest}. The meaning of a Q value and more details on this test is provided by \citet{kawaler88}. Basically, this test provides the most common values of period spacings that exist in the data. The result of our mode identification based on the asymptotic period spacing is also presented in \'echelle diagrams, which we discuss in Section\,\ref{echelle}.

\begin{figure*}
\includegraphics[width=\textwidth]{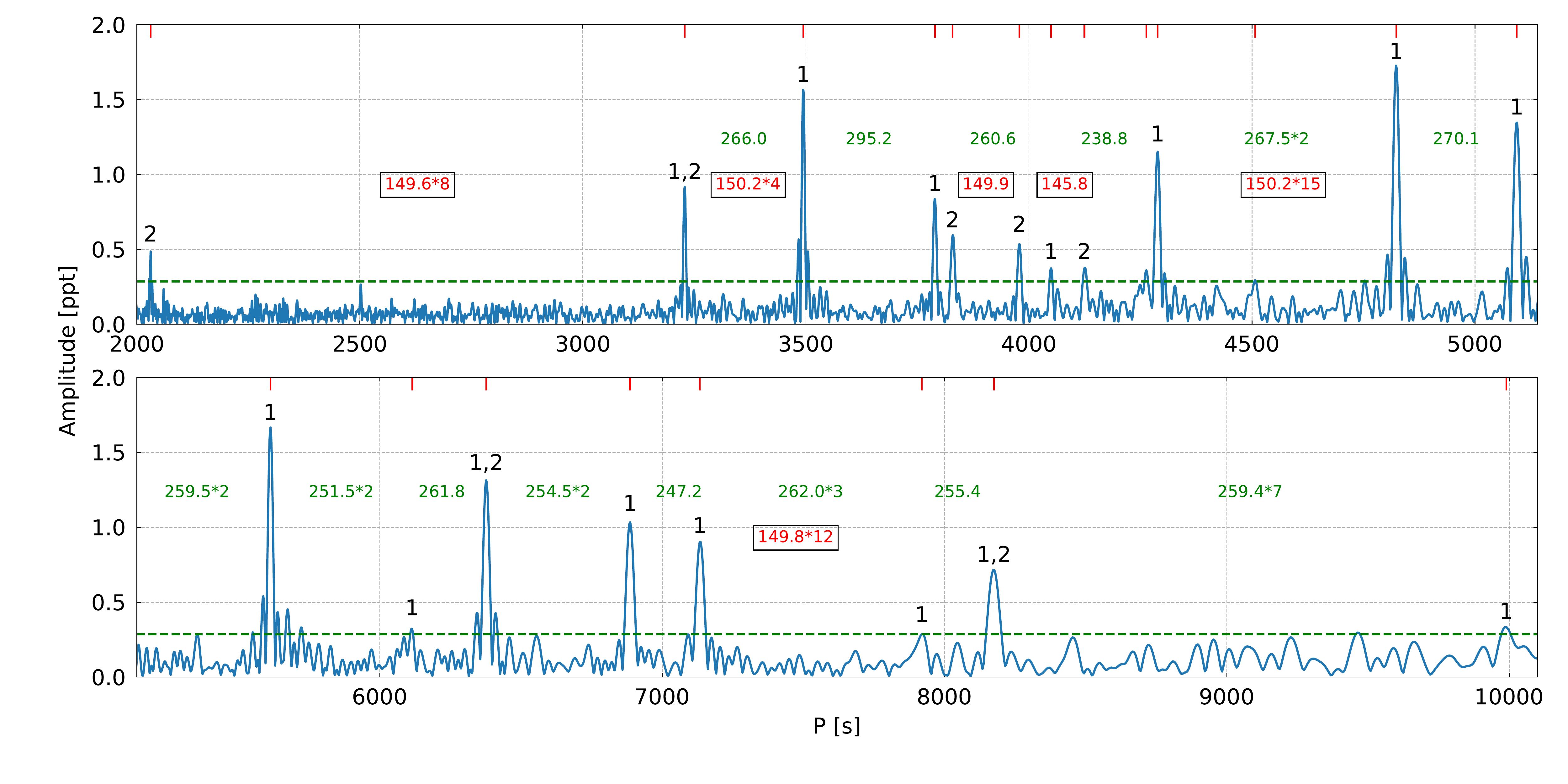}
\caption{A close-up of the amplitude spectrum of SB\,459 plotted in period instead of frequency. The green horizontal dashed line denotes the 4.5$\upsigma$ detection threshold. The values of modal degrees are shown on top of each identified modes. The values between frequencies denote period spacing between two overtones of the same degree (green for dipole and boxed red for quadrupole modes).}
\label{fig:067584818ft}
\end{figure*}

\textbf{SB\,459}\\
The KS test shows a common spacing of periods around 260\,s shown in the left panel of Figure\,\ref{fig:kstest}. We identified 12 dipole modes, four quadrupole modes and three peaks satisfying both sequences. We marked them in the amplitude spectrum in Figure\,\ref{fig:067584818ft}. Linear fits provide the average period spacings of 259.16(56)\,s and 149.89(5)\,s for dipole and quadrupole modes, respectively.

\begin{figure*}
\includegraphics[width=\textwidth]{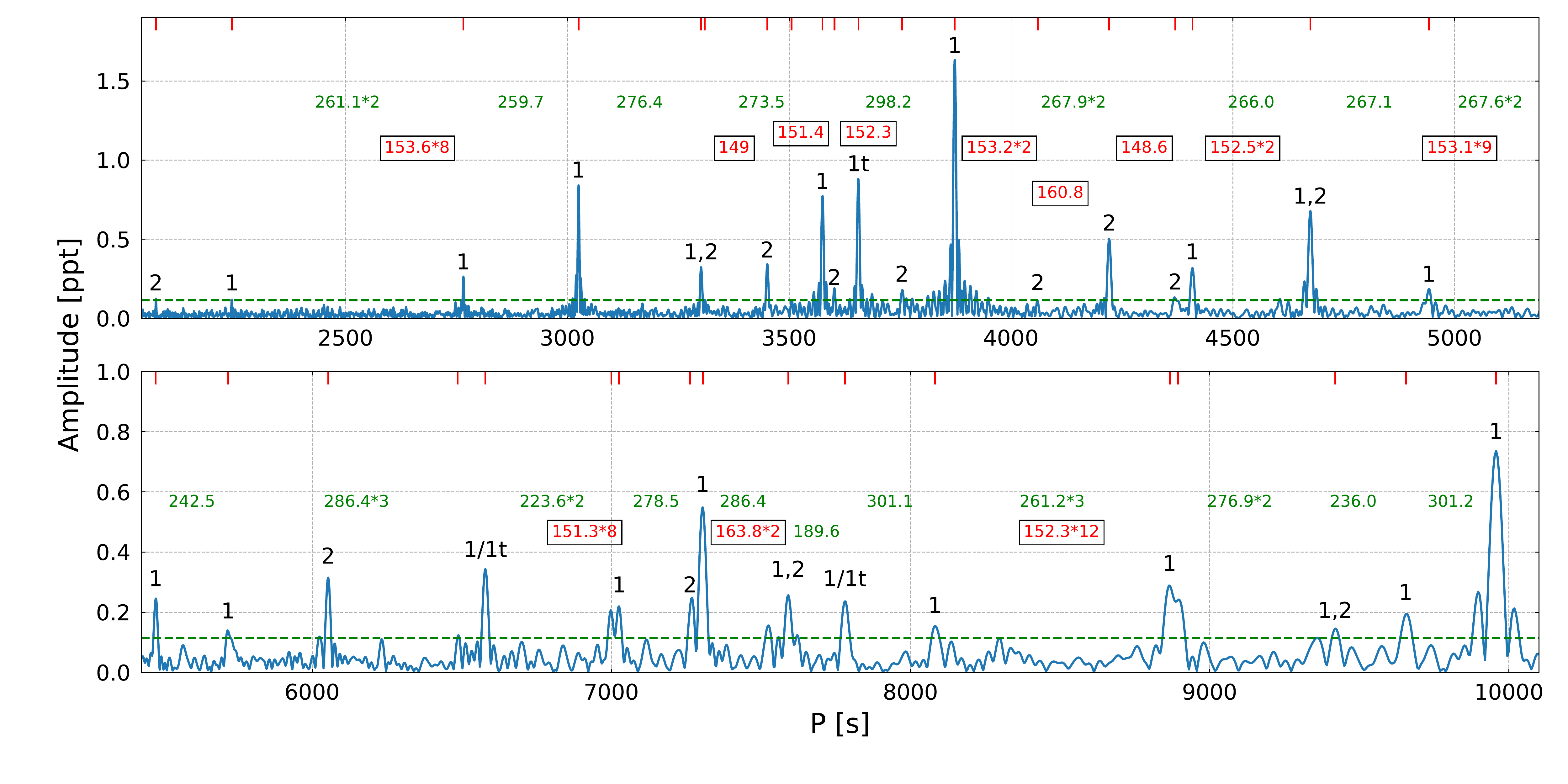}
\caption{Same as Figure\,\ref{fig:067584818ft} but for SB\,815.}
\label{fig:169285097ft}
\end{figure*}

\textbf{SB\,815}\\
In our analysis, we detected six frequencies in the p-mode region and we excluded those from our KS test, which eventually points at a common spacing of around 264\,s (middle panel in Figure\,\ref{fig:kstest}). We arrived at two possible solutions, which we present in Fig.\,\ref{fig:echelles} and \ref{fig:169285097ft}.\\
\textit{Solution\,1:}\\
In this solution, we identified 17 dipole modes, nine quadrupole modes, while four peaks fit both sequences. Linear fits provide the average period spacings of 265.04(73)\,s and 153.02(11)\,s for dipole and quadrupole modes, respectively. We identified a frequency 273.537\,$\upmu$Hz with an amplitude of 33.4$\upsigma$, where $\upsigma$ denotes an average noise level, which fits neither dipole nor quadrupole sequence. Its amplitude is also much higher to consider it to be any {\it l}\,$\geq$\,3 mode due to the surface cancellation effect \citep{dziembowski77}. Therefore, we assigned it as a trapped dipole mode. This mode identification looks fairly good, but two frequencies 128.523 and 151.999\,$\upmu$Hz differ excessively from the mean period spacing\,(28.5\% and 15.6\%, respectively). To justify these extreme deviations we followed the theoretical consideration provided by \citet{charpinet13} in Figure\,4, which presents that thin hydrogen envelope sdBVs show higher deviations from the mean period spacing.\\
\textit{Solution\,2:}\\
This solution considers those two extremely deviated frequencies as candidate for trapped modes. These peaks have moderate amplitudes (8.3 and 13.1\,$\upsigma$ respectively) and they do not fit the quadrupole sequence any better. Therefore, taking these two as trapped modes sorts out large deviations in the period sequence of the dipole modes. In this solution we are left with 15 dipole modes with average period spacing of 265.15(57)\,s and three trapped modes.

\begin{figure*}
\includegraphics[width=\textwidth]{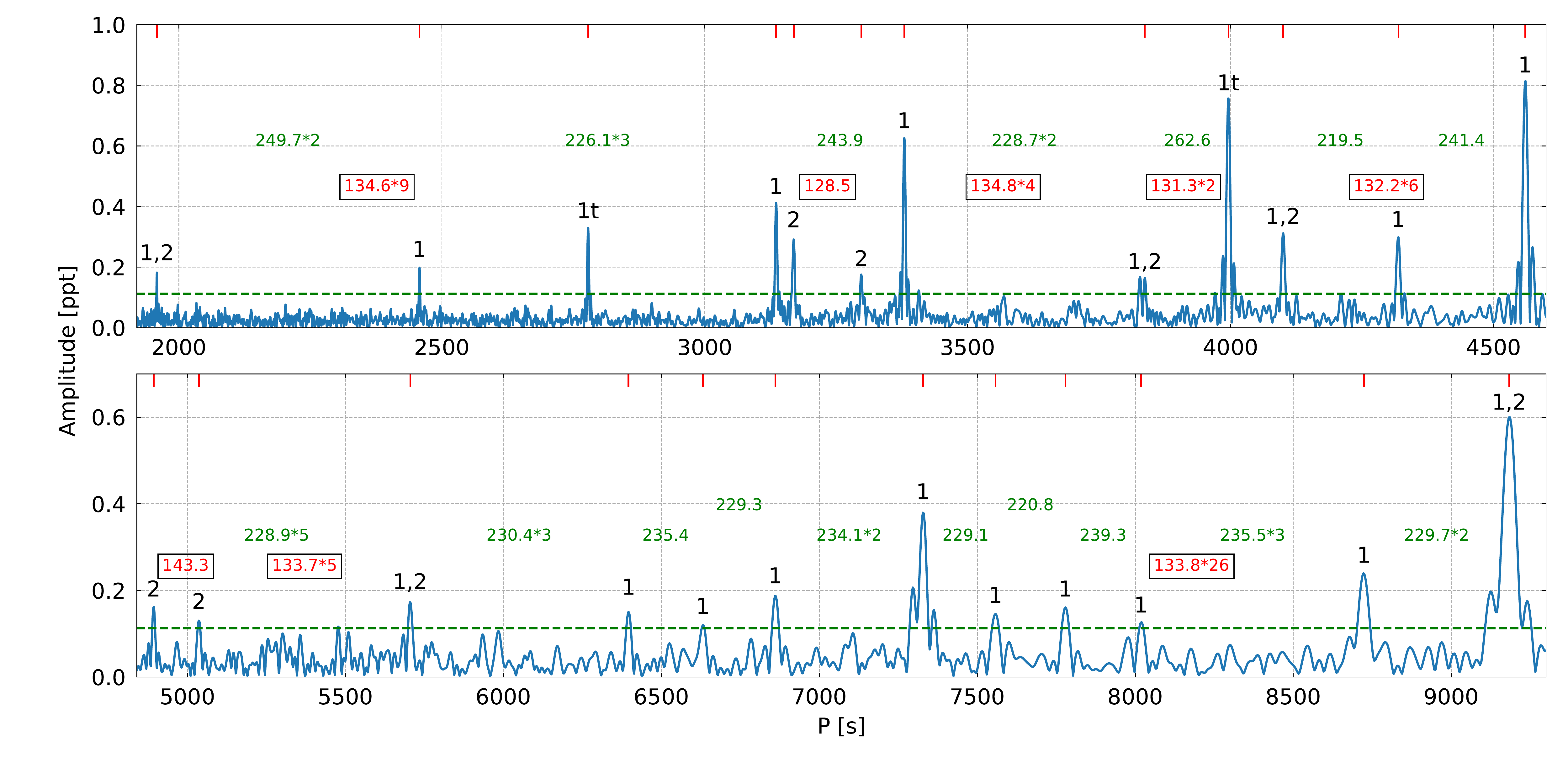}
\caption{Same as Figure\,\ref{fig:067584818ft} but for PG\,0342+026.}
\label{fig:457168745ft}
\end{figure*}

\textbf{PG\,0342+026}\\
The KS test points at a common spacing around 232\,s (right panel in Figure\,\ref{fig:kstest}). There is another minimum of a $\log{Q}$ value at 116\,sec. It is close to the expected value of a period spacing of quadrupole modes (132\,sec), however it is half the period spacing of dipole modes, which sometimes appears in this test. We identified 13 dipole modes, four quadrupole modes and five modes satisfying both sequences. We marked all identified modes in the amplitude spectrum in Figure\,\ref{fig:457168745ft}. A linear fit provides the average period spacings 232.25(30)\,s and 133.74(10)\,s for dipole and quadrupole modes, respectively. Two frequencies seem to be candidates for trapped modes and we refer to Section\,\ref{echelle} for more details.

\begin{figure*}
\includegraphics[width=\textwidth]{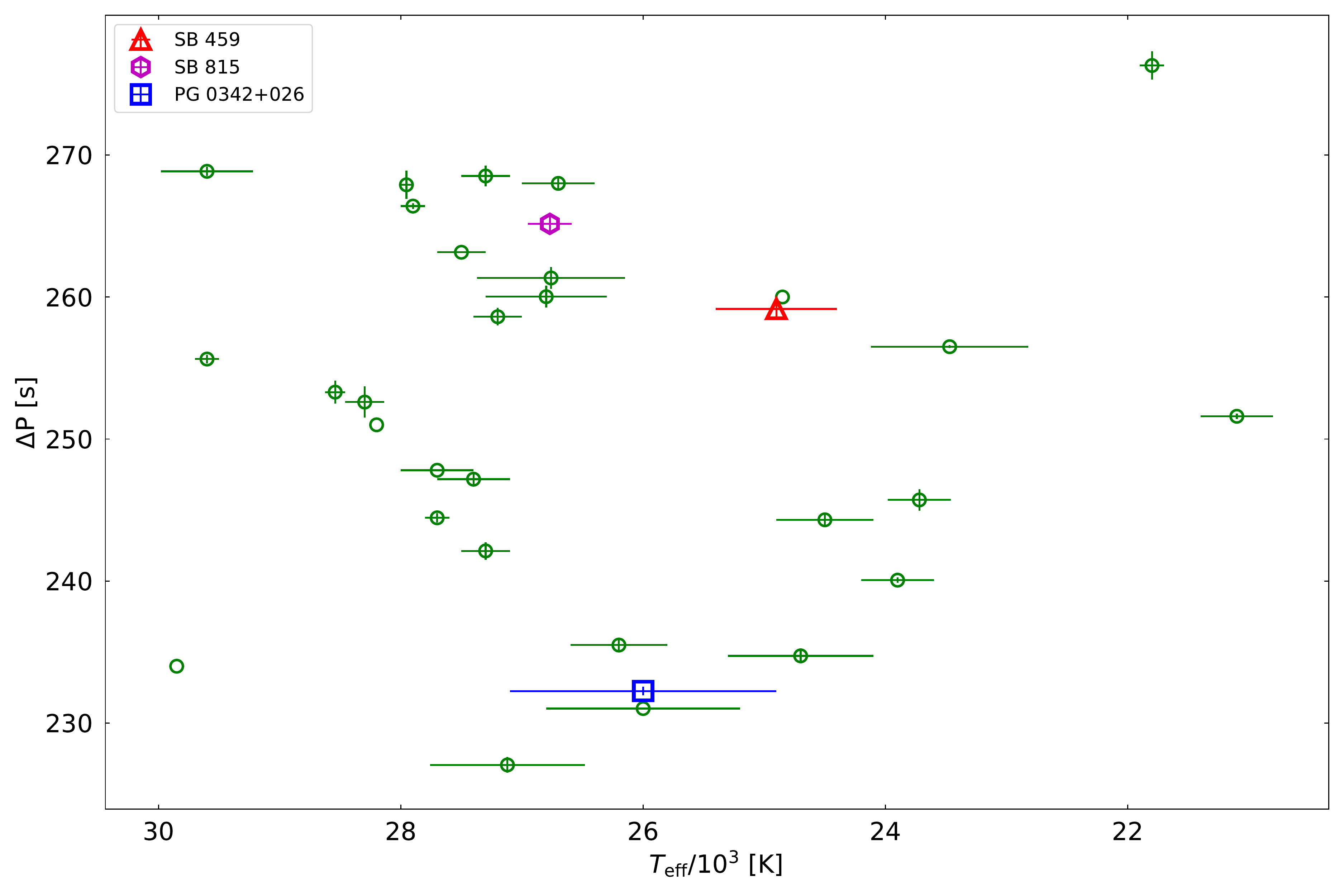}
\caption{Effective temperature in function of an average period spacing.}
\label{fig:tvsdp}
\end{figure*}

We have collected average period spacings ($\Delta$P) as a function of effective temperature for sdBV stars from the literature. All these collected information for 27 sdBVs is provided in Table\,\ref{tab:table5}. We plotted these two parameters in Figure\,\ref{fig:tvsdp}. We stress that the sample is not very large yet and any conclusion maybe biased. The first try of finding correlation between $\Delta$P and T$_{\rm eff}$ has been undertaken by \citet{reed11} with null result. We increased the number of points but our plot shows that still no clear correlation is present. There are zones of avoidance, though they may just be lacking data points as a consequence of a small sample. Therefore, based on our findings, we conclude that the average period spacing does not correlate with T$_{\rm eff}$ and so $\Delta$P does not translate to a specific T$_{\rm eff}$ and vice versa.

\subsection{\'Echelle diagrams and candidates for trapped modes}
\label{echelle}
\begin{figure*}
\includegraphics[width=\textwidth]{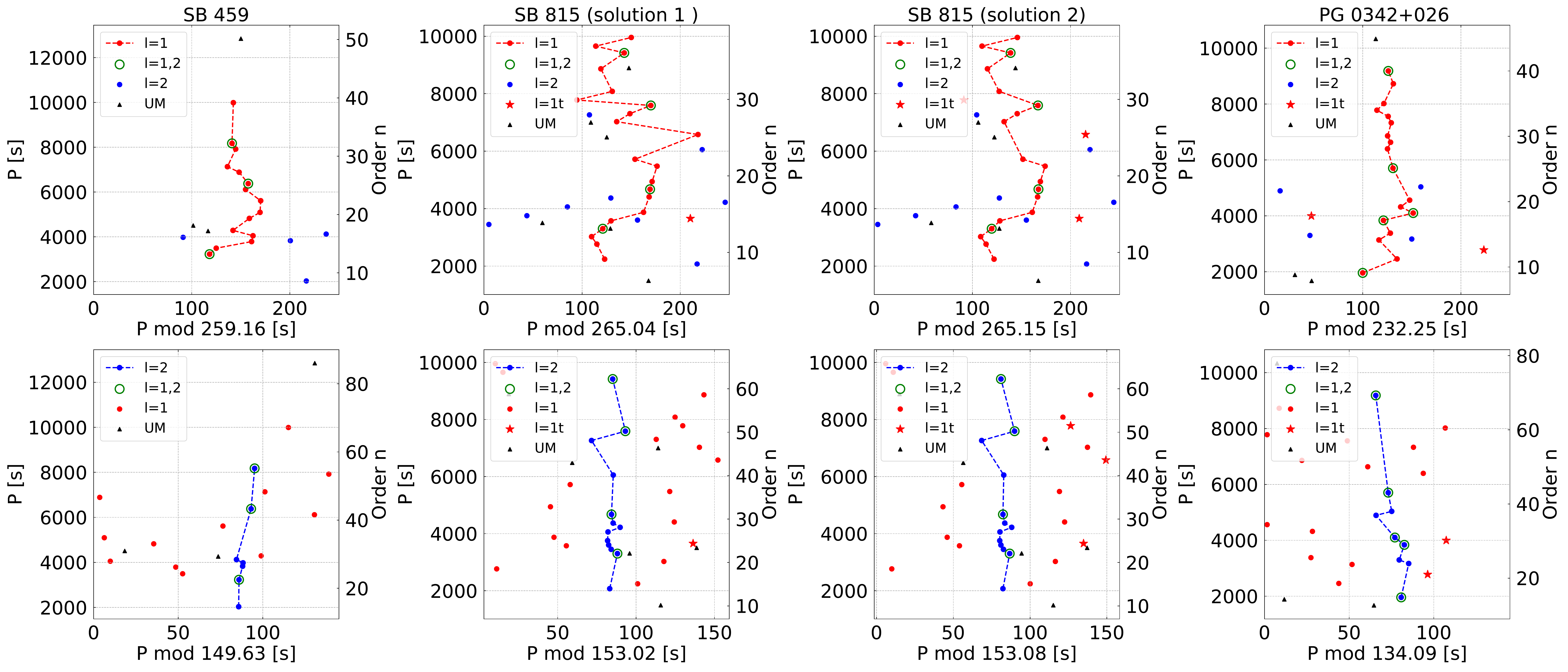}
\caption{\'Echelle diagrams for dipole (top panels) and quadrupole (bottom panels) modes. In legends, `l=1t' denotes dipole trapped modes and `UM' denotes undefined modes.}
\label{fig:echelles}
\end{figure*}

The \'echelle diagrams are very useful tools for testing the identification of the modes by means of the asymptotic period spacing. These diagrams represent P\,{\it mod}\,$\Delta$P in function of P, where P is the pulsation period and $\Delta$P is a period spacing. We present the diagrams for all three targets in Figure\,\ref{fig:echelles}. For SB\,815 we include two solutions. The upper panels show the \'echelle diagrams for dipole modes while the bottom panels show the diagrams for quadrupole modes. Peaks satisfying both the sequences have been added to both dipole and quadrupole \'echelle diagrams and represented with green color points. The right vertical axes show the radial orders with respect to an offset $n_l$ from the real radial {\it k} order. The {\it k} number can only be determined from modeling \citep[e.g.][]{charpinet00}.

The asymptotic relation, defined by equation\,\ref{eq:2}, is strict only for homogeneous stars. In that case, standing waves of g-modes oscillate in a cavity created {\it e.g.} by the convective zone and the surface of a star. Then, the consecutive overtones are spaced equally in period and in an \'echelle diagram we can see a vertical ridge for a given modal degree. However, in a real star, as the density is not uniform, the ridge almost never becomes purely vertical. Some jitter appears. Since this feature is a consequence of a non uniform structure of a star, deviations from a vertical ridge bear information about a chemical profile and hence cavities. \citet{baran12b} reported on a deviation from the vertical ridge, being a common property of many sdB stars \citep{baran19}. The so-called a "hook" feature has never been explained thus far but surely must be accounted for if reliable models are to be calculated.

In a few cases, frequencies did not fit well either sequence. The reason may be hidden deep in the sdB interior where the H/He or C/He transition layers between the convective core and the surface appear. These boundaries may contribute to create additional cavities causing some modes to be imprisoned in smaller cavities. Those modes are called trapped modes and they do not follow an asymptotic sequence. The theoretical explanation was provided by \citet{charpinet00} and \citet{ghasemi17}.

There is a "hook" feature in SB\,459 between 3\,000 and 7\,000\,s, while in PG\,0342+026 the feature is not as pronounced. The largest jitter appears in SB\,815 which deviates from the mean period spacing by 28.5\%. The upper part of the ridge is not smooth, winding from side to side. That is why we decided to present two solutions for this target. In the absence of multiplets, it is always difficult to make sure that a mode identification is fully correct. Our first solution contains the largest jitter but it provides the "hook" feature in between 3\,000 and 7\,000\,s. In our second solution, we removed two extremely deviated points (6579.0\,s and 7780.7\,s) from the dipole mode sequence, and marked them as trapped mode candidates. In the latter solution the \'echelle diagram looks more smooth and still shows the "hook" feature. With no multiplets detected our identification will always suffer from doubts in modal degree assignment, mostly because period spacing sequences of different modal degree cross each other and some of the modes are fitting both sequences fairly well. In case of high amplitude frequencies we prefer {\it l}\,=\,1 rather than higher degrees. The ridges of quadrupole modes are fairly short and those modes are mostly leftovers from {\it l}\,=\,1 assignment.

One of the best tools to look for trapped modes is a reduced period diagram. The diagram presents a reduced period $\Pi={\rm P} \cdot \sqrt{l(l+1)}$ in function of a reduced period spacing $\Delta\Pi=\Delta {\rm P} \cdot \sqrt{l(l+1)}$. This multiplication causes sequences of all modal degrees to overlap. Overall, the shape of the plot would be similar to what we see in the \'echelle diagrams, though it will be twisted, so the ridge is now horizontal. Modes with different modal degrees overlap, however, what is more important, the candidates for trapped modes of different degrees also overlap. It can be clearly seen in the papers by e.g. \citet{ostensen14,uzundag17,baran17}. The actual periods of those trapped modes differ between modal degrees, so it is not easy to spot them in amplitude spectra, however the multiplicative factor brings them all in one place in this diagram.

\begin{figure*}
\includegraphics[width=\textwidth]{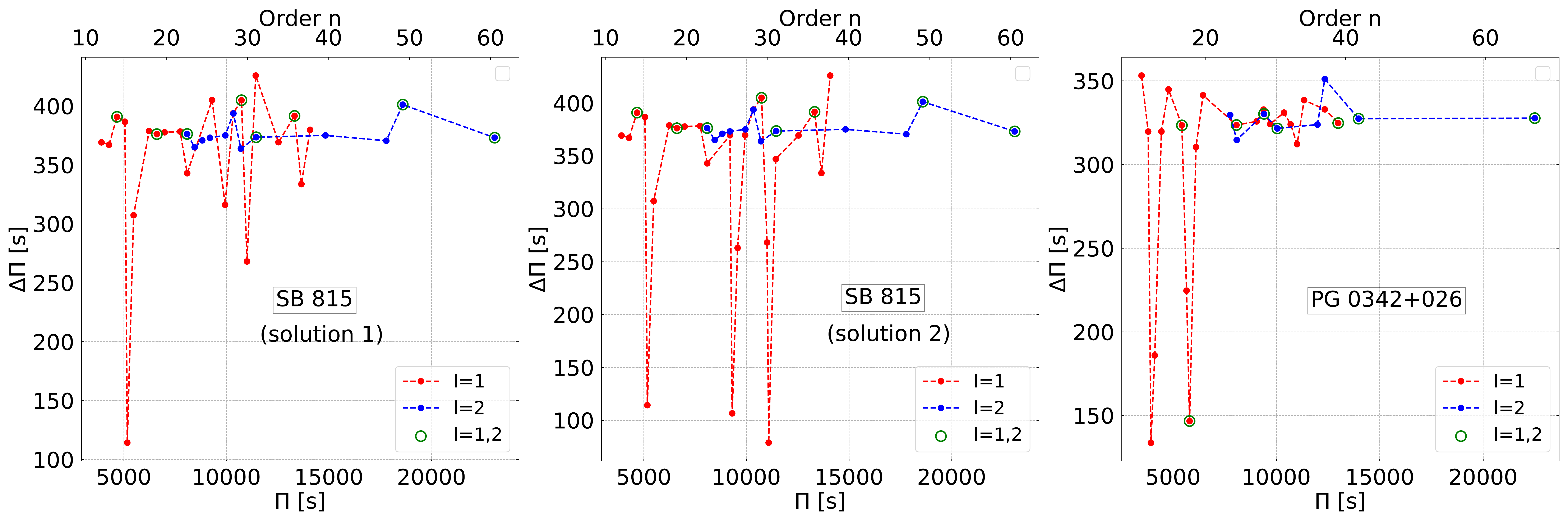}
\caption{Reduced period diagrams for SB\,815 and PG\,0342+026. See text for more details.}
\label{fig:rpd}
\end{figure*}

We show the reduce period diagrams for two targets, SB\,815 (two solutions) and PG\,0342+026 in Figure\,\ref{fig:rpd}. In SB\,459, the sequence of quadrupole modes is too short, not pointing at any trapped mode candidates, which makes the diagram completely inconclusive, and that is why we decided not to present it. In the first solution of SB\,815 and in PG\,0342+026, the candidates for trapped modes appear to be at the shortest periods. It looks similar to the diagrams reported by the other authors mentioned above. In SB\,815 we find either one (solution 1) or three (solution 2), while in PG\,0342+026 we find two candidates for trapped modes. Two longest periods trapped modes in SB\,815 (solution 2) and two trapped modes in PG\,0342+026 are separated by almost 2\,000\,sec. It agrees with values reported by the other authors and calculated from theoretical considerations reported by \citet{charpinet00}. Unluckily, in PG\,0342+026 the quadrupole sequence do not extend to overlap with those candidates and therefore we cannot confirm trapped mode identification. Likewise in SB\,815 (solution 1). In the case of solution 2, although the dipole and quadrupole sequences overlap, we detected no quadrupole trapped modes candidates. This makes those dipole trapped modes candidates less reliable. They can still serve as an additional constraint in modeling, help deriving the most reliable solution and understand the chemical profile inside sdB stars, which is responsible for trapped modes.

\section{Evolutionary status}
\begin{figure*}
\includegraphics[width=\textwidth]{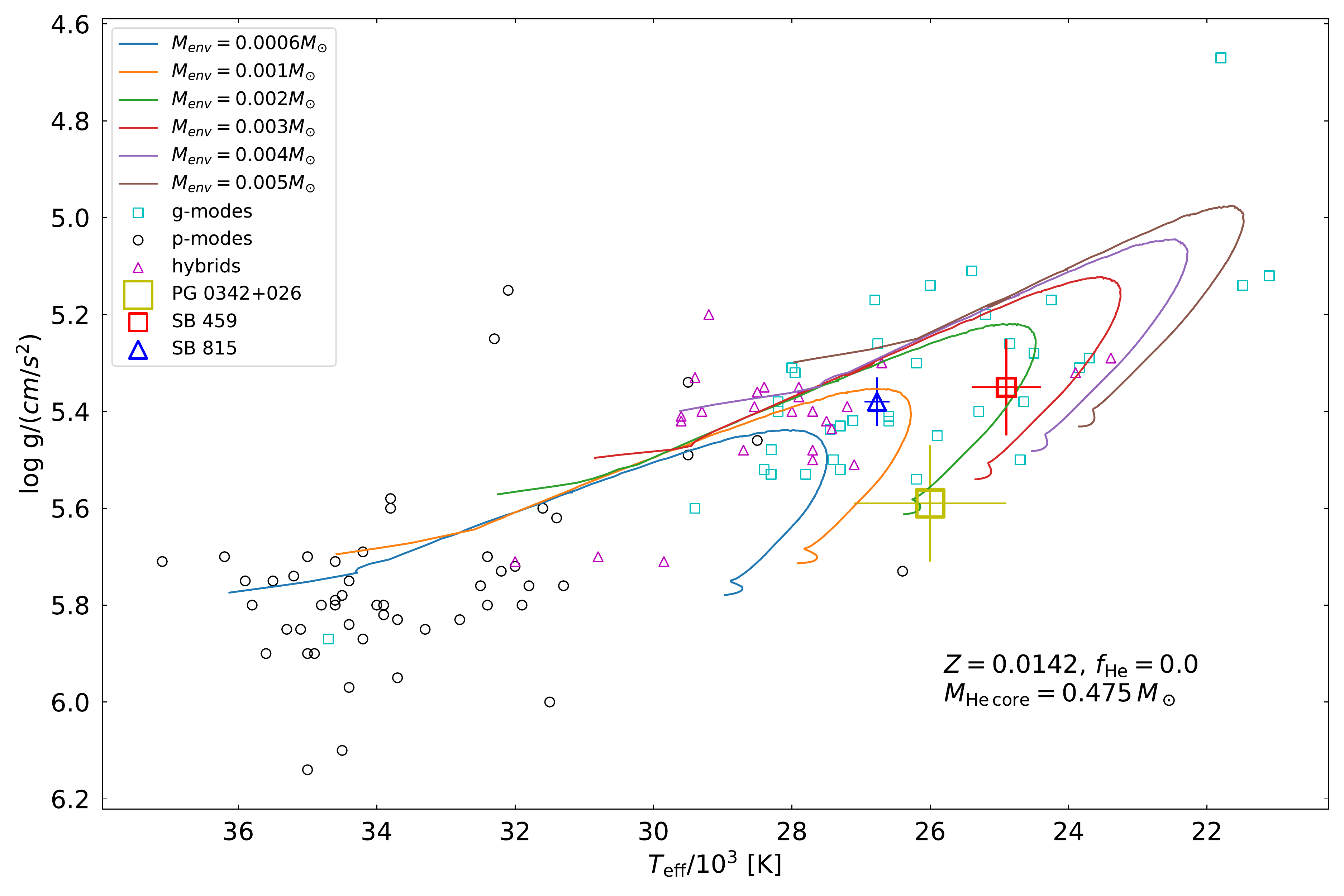}
\caption{Evolutionary tracks of sdB stars in the $\log{g}$ - $T_\mathrm{eff}$ diagram for sdB stars with initial mass of $M_\mathrm{i} = 1.0\,M_\odot$, mass of helium core $M_\mathrm{He,\,core} = 0.475\,M_\odot$ and envelope masses of $6\times 10^{-4} - 5\times 10^{-3}$\,M$_\odot$. Observational data are taken from \citet{holdsworth17} along with three sdBVs explained in this paper.}
\label{fig:tvsg}
\end{figure*}

The stellar atmospheric parameters such as effective temperature T$_{\rm eff}$ and surface gravity $\log{g/(\rm{cm\,s^{-2}}}$) have great importance to determine physical conditions of stellar atmospheres. We have taken these two spectroscopic parameters of all sdBVs known to date from \citet{holdsworth17} and for our three TESS targets from Table\,\ref{tab:table1}. We have plotted these three targets along with 118 other previously known sdBVs in the effective temperature - surface gravity diagram (Figure\,\ref{fig:tvsg}). In the plot we can distinguish three different regions i.e. low T$_{\rm eff}$ and $\log{g/(\rm{cm\,s^{-2}}}$) containing g-mode pulsators (shown in cyan squares), high T$_{\rm eff}$ and $\log{g/(\rm{cm\,s^{-2}}}$) containing p-mode pulsators (shown in black circles), and the hybrid pulsators region (shown in magenta triangles) containing pulsators that show both p- and g-modes. Three TESS targets have been shown with bigger symbols along with the error bars. SB\,459 and PG\,0342+026 are located among g-mode pulsators, which is consistent with the frequency content of these two stars. The amplitude spectrum of SB\,815 contains both g-mode and p-mode, which is also confirmed by its location in the plot.

\begin{table}
\centering
\caption{Effective temperature and average period spacing data for known sdBVs. The first reference is for $\Delta$\,P and the second is for T$_{\rm eff}$. References: 1.\,\citet{reed18b}, 2.\,\citet{silvotti19}, 3.\,\citet{reed11}, 4.\,\citet{holdsworth17}, 5.\,Sanjayan et al.\,(in preparation), 6.\,\citet{charpinet19}, 7.\,\citet{reed20}\,(submitted)}
\label{tab:table5}
\begin{tabular}{cccc}
\hline
Name & $\Delta$P\,[s] & T$_{\rm eff}$\,[kK] & References\\
\hline
KIC\,1718290 & 276(1) & 21.8(1) & 1,4\\
KIC\,2437937 & 234.73(52) & 24.7(6) & 5,4\\
KIC\,2438324 & 235.49(51) & 26.2(4) & 5,4\\
KIC\,2569576 & 244.31(46) & 24.5(4) & 5,4\\
KIC\,2697388 & 240.06(19) & 23.9(3) & 1,4\\
KIC\,2991403 & 268.52(74) & 27.3(2) & 1,4\\
KIC\,3527751 & 266.4(2) & 27.9(1) & 1,4\\
KIC\,5807616 & 242.12(62) & 27.3(2) & 1,4\\
KIC\,7664467 & 260.02(77) & 26.8(5) & 1,4\\
KIC\,7668647 & 247.8 & 27.7(3) & 1,4\\
KIC\,8302197 & 258.61(62) & 27.2(2) & 1,4\\
KIC\,9472174 & 255.63(30) & 29.6(1) & 1,4\\
KIC\,10001893 & 268.0(5) & 26.7(3) & 1,4\\
KIC\,10553698 & 263.15 & 27.5(2) & 1,4\\
KIC\,10670103 & 251.6(2) & 21.1(3) & 1,4\\
KIC\,11179657 & 231.02(2) & 26.0(8) & 1,4\\
KIC\,11558725 & 244.45(32) & 27.7(1) & 1,4\\
EPIC\,201206621 & 268(1) & 27.954(54) & 1,4\\
EPIC\,202065500 & 234 & 29.85 & 1,4\\
EPIC\,203948264 & 261.34(78) & 26.76(61) & 1,4\\
EPIC\,211696659 & 227.05(56) & 27.12(64) & 1,4\\
EPIC\,211779126 & 253.3(8) & 28.542(82) & 1,4\\
EPIC\,212707862 & 252.6(1.1) & 28.298(162) & 1,4\\
EPIC\,218366972 & 251 & 28.2 & 1,4\\
EPIC\,218717602 & 260 & 24.85 & 1,4\\
EPIC\,220641886 & 256.5(1) & 23.47(65) & 2,2\\
KPD\,0629-0016 & 247.17(48) & 27.4(3) & 3,4\\
TIC\,013145616 & 268.85(32) & 29.60(38) & 7,7\\
TIC\,278659026 & 245.71(75) & 23.72(26) & 6,6\\
\hline
\end{tabular}
\end{table}

In Fig.\,\ref{fig:tvsg} we also plotted theoretical evolutionary tracks to assess the evolutionary status of our three targets. The tracks have been calculated using publicly available open source code {\texttt MESA} \citep[Modules for Experiments in Stellar Astrophysics;][]{paxton11,paxton13,paxton15,paxton18,paxton19}, version 11701. We started with a pre-main-sequence model of a solar mass star, assumed a proto-solar chemical composition of \citet{asplund09} (Z\,=\,0.142, Y\,=\,0.2703), and evolved the model to the tip of the red giant branch. Then, before the helium flash, we removed most of its mass leaving only a residual hydrogen envelope on top of the helium core. The model was then relaxed to an equilibrium state and evolved until the depletion of helium in the core. All physical and numerical details of the models are discussed in Ostrowski\,et\,al.\,(in preparation). The models use predictive mixing to ensure proper growth of the convective core during the course of evolution \citep{paxton18}.  The evolutionary tracks presented in Fig.\,\ref{fig:tvsg} show stable core He burning phase of the sdB evolution. Different tracks correspond to models with different hydrogen envelope masses ($M_{\rm env}$\,=\,6\,$\times$\,10$^{-4}$\,--\,5\,$\times$\,10$^{-3}$\,$M_\odot$). It may be noted that the effect of increasingly more massive hydrogen envelopes is to shift the evolutionary tracks towards lower effective temperatures.


The sdBs start their evolution toward lower effective temperatures and lower surface gravities. The direction of the evolution is reversed when the central helium abundance drops below about 10\%. The presented tracks fit the location of all our three targets very well and firmly confirm the three stars to be sdBs. All three targets are located on the He-core burning tracks. SB\,459 fits really well to a track with an envelope mass of M$_{\rm env}$\,=\,2\,$\times$\,10$^{-3}$M$_\odot$ and still has more than half of its initial helium abundance available in the core. SB\,815 is more advanced in its evolution with a central helium abundance of about ten percent and it is better fitted by a track with an envelope mass of M$_{\rm env}$\,=\,1\,$\times$\,10$^{-3}$M$_\odot$. The spectroscopic parameters of the star are determined with better precision than those of other two targets. PG\,0342+026 seems to be the youngest sdBVs among the three stars, at the beginning of the sdB phase. The envelope mass of the star may vary between M$_{\rm env}$\,=\,1\,--\,3\,$\times$\,10$^{-3}$M$_\odot$.

\section{Summary}
In this paper we report our asteroseismic analysis of three sdBV stars observed by the TESS satellite. We have analyzed amplitude spectra to detect pulsation modes and we used the asymptotic period spacing to describe modes' geometries. For SB\,459 we found 12 dipole modes, four quadrupole modes and three modes that can be assigned with either modal degree. For SB\,815 we did not find a unique solution. In solution\,1 we identified 17 dipole modes, nine quadrupole modes and four modes that can be assigned with either modal degree. In solution\,2 we identified the same number of modes, however two dipole modes are considered candidates for trapped modes. In PG\,0342+026 we identified 13 dipole modes, four quadrupole modes and five modes that can be assigned with either modal degree. We found none multiplets and therefore our mode identification should be taken with caution.

The average period spacings of dipole modes is around 259\,s and 265\,s and 232\,s for SB\,459, SB\,815 and PG\,0342+026, respectively. In all three targets we detected only few quadrupole modes and hence average period spacing values for quadrupole modes calculated from the linear fits are not too precise. We used a theoretical relation between period spacings of dipole and quadrupole modes, instead.

We also found a few candidates for trapped modes, one/three in SB\,815 and two in PG\,0342+026. In the reduced period diagrams the trapped mode candidates are spaced by around 2000\,sec. This spacing is predicted by theoretical calculations and makes our conclusion more reliable, yet not absolutely convincing, since we detected no quadrupole trapped modes counterparts.

By making use of the high precision Gaia parallaxes and spectral energy distributions from the ultraviolet to the infrared spectral range we derived the fundamental stellar parameters mass, radius, and luminosity from spectroscopically determined effective temperatures and gravities. The results are consistent with the predictions of canonical stellar evolutionary models \citep{1993ApJ...419..596D}, however, with large uncertainties on stellar mass due to large uncertainties on $\log g$.

The location of our three sdBVs in the effective temperature - surface gravity diagram confirms that SB\,459 and PG\,0342+026 are g-mode dominated sdBVs and SB\,815 is g-mode dominated hybrid pulsator. Theoretical evolutionary tracks provide a coarse-grained approximation of physical properties of these stars like He-core and hydrogen envelope masses, sizes of their cores along with their evolutionary sdB stages. These tracks show that all three stars are during core-helium-burning phase, where SB\,815 is much more evolved than other two and PG\,0342+026 has just entered the sdB phase.


We also tried to look for any correlation between $\Delta$P and T$_{\rm{eff}}$ with all previously known g-mode sdBVs along with our three TESS targets. We found no correlations though. We suspect to see some correlations with increasing data points. The asteroseismic analysis of these targets will help to constrain models for these stars. This paper is our first attempt to list g-mode rich sdBVs observed in TESS and to do mode identifications for these targets.

\section*{Acknowledgements}
Financial support from the Polish National Science Center under projects No.\,UMO-2017/26/E/ST9/00703 and UMO-2017/25/B ST9/02218 is acknowledged. R.R., U.H. and A.I. gratefully acknowledge financial support by the Deutsche Forschungsgemeinschaft through grants HE1356,71-1 and IR190/1-1. Theoretical calculations have been carried out using resources provided by Wroclaw Centre for Networking and Supercomputing (http://wcss.pl), grant No. 265. Based on observations obtained at the European Organisation for Astronomical Research in the Southern Hemisphere under ESO observing program 0103.D-0511. Based on observations obtained at Las Campanas Observatory under the run code 0KJ21U8U. M.U. acknowledges financial support from CONICYT Doctorado Nacional in the form of grant number No: 21190886. RR has received funding from the postdoctoral fellowship programme Beatriu de Pin\'os, funded by the Secretary of Universities and Research (Government of Catalonia) and by the Horizon 2020 programme of research and innovation of the European Union under the Maria Sk\l{}odowska-Curie grant agreement No 801370. KJB is supported by the National Science Foundation under Award No.\,AST-1903828. WZ acknowledges the support from the Beijing Natural Science Foundation (No. 1194023) and the National Natural Science Foundation of China (NSFC) through the grant 11903005. This work has made use of data from the European Space Agency (ESA) mission {\it Gaia} (\url{https://www.cosmos.esa.int/gaia}), processed by the {\it Gaia} Data Processing and Analysis Consortium (DPAC, \url{https://www.cosmos.esa.int/web/gaia/dpac/consortium}). Funding for the DPAC has been provided by national institutions, in particular the institutions participating in the {\it Gaia} Multilateral Agreement. This publication makes use of data products from the Wide-field Infrared Survey Explorer, which is a joint project of the University of California, Los Angeles, and the Jet Propulsion Laboratory/California Institute of Technology, funded by the National Aeronautics and Space Administration. The national facility capability for SkyMapper has been funded through ARC LIEF grant LE130100104 from the Australian Research Council, awarded to the University of Sydney, the Australian National University, Swinburne University of Technology, the University of Queensland, the University of Western Australia, the University of Melbourne, Curtin University of Technology, Monash University and the Australian Astronomical Observatory. SkyMapper is owned and operated by The Australian National University's Research School of Astronomy and Astrophysics. The survey data were processed and provided by the SkyMapper Team at ANU. The SkyMapper node of the All-Sky Virtual Observatory (ASVO) is hosted at the National Computational Infrastructure (NCI). Development and support the SkyMapper node of the ASVO has been funded in part by Astronomy Australia Limited (AAL) and the Australian Government through the Commonwealth's Education Investment Fund (EIF) and National Collaborative Research Infrastructure Strategy (NCRIS), particularly the National eResearch Collaboration Tools and Resources (NeCTAR) and the Australian National Data Service Projects (ANDS).




\bibliographystyle{mnras}
\bibliography{bibliography.bib} 








\bsp	
\label{lastpage}
\end{document}